\documentclass[aps,showpacs,preprintnumbers]{revtex4}
\usepackage{graphicx}
\usepackage{epsfig}
\usepackage{color}
\usepackage{amsmath}
\usepackage{hyperref}
\usepackage{epsf,amssymb,amsfonts,amsmath}
\graphicspath{{figs/}}

\begin{document}
\title{Gravitational Waves and Primordial Black Hole Productions from Gluodynamics by Holography}

\author{Song He$^{1,2}$}
\email{hesong@jlu.edu.cn}

\author{Li Li$^{3,4,5}$}
\email{liliphy@itp.ac.cn}

\author{Zhibin Li$^{6}$}
\email{lizhibin@zzu.edu.cn}

\author{Shao-Jiang Wang$^{3}$}
\email{schwang@itp.ac.cn}

\affiliation{ $^{1}$ Center for Theoretical Physics and College of Physics, Jilin University,
	Changchun 130012, China}
\affiliation{ $^{2}$ Max Planck Institute for Gravitational Physics (Albert Einstein Institute), Am Muhlenberg 1, 14476 Golm, Germany}
\affiliation{ $^{3}$ CAS Key Laboratory of Theoretical Physics, Institute of Theoretical Physics,
	Chinese Academy of Sciences, Beijing 100190, China}
\affiliation{ $^{4}$ School of Fundamental Physics and Mathematical Sciences,
	Hangzhou Institute for Advanced Study, University of Chinese Academy of Sciences, Hangzhou 310024, China}
\affiliation{ $^{5}$ Peng Huanwu Collaborative Center for Research and Education, Beihang University, Beijing 100191, China}
\affiliation{ $^{6}$ School of Physics and Microelectronics, Zhengzhou University, Zhengzhou 450001, China}


\begin{abstract}
Understanding the nature of quantum chromodynamics (QCD) matter is important but challenging due to the presence of non-perturbative dynamics under extreme conditions. We construct a holographic model describing the gluon sector of QCD at finite temperatures in the non-perturbative regime. The equation of state as a function of temperature is in good accordance with the lattice QCD data. Moreover, the Polyakov loop and the gluon condensation, which are proper order parameters to capture the deconfinement phase transition, also agree quantitatively well with the lattice QCD data. We obtain a strong first-order confinement/deconfinement phase transition at $T_c=276.5\,\text{MeV}$ that is consistent with the lattice QCD prediction. Based on our model for a pure gluon hidden sector, we compute the stochastic gravitational waves and primordial black hole (PBH) productions from this confinement/deconfinement phase transition in the early Universe. The resulting stochastic gravitational-wave backgrounds are found to be within detectability in the International Pulsar Timing Array and Square Kilometre Array in the near future when the associated productions of PBHs saturate the current observational bounds on the PBH abundances from the LIGO-Virgo-Collaboration O3 data.
\end{abstract}
		

\maketitle

	
	
\section{Introduction}
The early Universe before the big bang nucleosynthesis is opaque to electromagnetic waves. Thanks to the recent gravitational-wave detections, future observations of stochastic gravitational wave backgrounds (SGWBs) would reveal the new physics~\cite{Cai:2017cbj,Caprini:2019egz, Hindmarsh:2020hop,Bian:2021ini} from the early Universe, including various first-order phase transitions (FOPTs) beyond the standard model of particle physics (see~\cite{Cai:2022bcf} and references therein for a model summary). It was recently found that the FOPT not only associates with SGWBs but also produces primordial black holes (PBHs) in general~\cite{Liu:2021svg,Liu:2022lvz} (see also~\cite{Hashino:2021qoq} for an explicit example from the electroweak phase transition), regardless of the specific particle physics model for realizing the FOPTs (see also~\cite{Baker:2021nyl,Baker:2021sno,Kawana:2021tde,Huang:2022him,Marfatia:2021hcp} for other specific mechanisms of PBH productions during some particular kinds of FOPTs). In particular, for the FOPT around the QCD scale, the associated SGWBs can be probed by the Pulsar Timing Array (PTA) and Square Kilometre Array (SKA) observations, and the associated PBH abundance could be constrained by the  LIGO-Virgo-Collaboration (LVC) network. While the QCD phase transition in the standard model at small baryon chemical potentials is cross-over, the pure gluon case features a confinement FOPT. This is a minimal scenario among many extensions of the standard model and is ideal as a benchmark model. Therefore, we will study pure gluons in this work for a realization of the FOPT around the QCD scale with associated productions of SGWBs and PBHs. Note here that the large density perturbations required to form PBHs from FOPTs are generated during FOPTs in the radiation era. This is totally different from other popular PBH production mechanisms with the large density perturbations induced from the large curvature perturbations originated from the inflationary period, for example, in the non-minimal curvaton models~\cite{Pi:2021dft,Meng:2022low}.
	
On the other hand, investigating the pure gluon system is important to understand the nature of hot and dense QCD matter formed in the early Universe and the laboratory. In particular, the gluon dynamics is dominant during $10^{-5}$ seconds into the expansion of the early Universe ~\cite{Shuryak:1992wc,McLerran:1993ni,Alam:1994sc,LHCb:2015yax} and an extremely rapid thermalization~\cite{Heinz:2002gs,Xu:2004mz,Broniowski:2008qk} in nucleus-nucleus collisions. On the theoretical side, the thermodynamics of the pure-gauge sector can be relevant to capture the essential qualitative features of the deconfinement, which is characterized by center symmetry and shows all the infrared difficulties of QCD. Due to the strong coupling, non-perturbative approaches are necessary for quantitative studies of its dynamics.
In addition to the lattice QCD that relies on massive computing power, an alternative non-perturbative approach is to employ the gauge/gravity correspondence~\cite{Maldacena:1997re,Gubser:1998bc,Witten:1998qj} that provides a powerful way to study strongly coupled non-Abelian gauge theories (see also~\cite{Gubser:2008yx,Gursoy:2008bu,Panero:2009tv,Jarvinen:2021jbd,Zhu:2021vkj} for earlier studies on the pure gluon system from holography).

In this work, we provide a bottom-up holographic QCD model for the pure gluon QCD system. The equation of state (EoS) quantitatively matches the pure gluon system in lattice QCD~\cite{Boyd:1996bx,Caselle:2018kap}. The confinement phase transition in gauge theory is characterized by the Polyakov loop operator $\langle \mathcal{P} \rangle$ which is finite in the deconfined phase and becomes vanishing in the confined phase for pure gluon~\cite{Kuti:1980gh,McLerran:1981pb}. The temperature dependence of $\langle \mathcal{P} \rangle$ from our model matches the lattice simulation~\cite{Gupta:2007ax} perfectly, and the predicted critical temperature $T_c=276.5\, \text{MeV}$ agrees with the expectation in the literature~\cite{Boyd:1996bx,Borsanyi:2012ve}. Moreover, another important quantity characterizing the deconfinement phase transition in a pure gluon system is the gluon condensation, which can be computed to be quantitatively consistent with the trace anomaly~\cite{Boyd:1996bx}. The strong FOPT in the early Universe is also a potentially important source for the production of SGWBs and PBHs. Our present model provides a reliable scenario for generating gravitational waves from a FOPT of a pure $SU(3)$ Yang-Mills sector. The resulting gravitational wave signals could be detected in the upcoming International PTA (IPTA) and SKA observations for the associated PBH abundance saturating the current observational bounds from the LVC constraints.
	

\section{Model}
We now build up a holographic model for the $SU(3)$ pure gluon system with the action of the following form.
	
\vspace{-5mm}
\begin{align}\label{action1}
S=\frac{1}{2\kappa_N^2}\int \mathrm{d}^{5}x \sqrt{-g} \left[\mathcal{R}-\frac{1}{2}\nabla_\mu \phi \nabla^\mu \phi-V(\phi)\right]
\end{align}
with the minimal cost of degrees of freedom to capture the essential dynamics. The gravitational theory includes only two fields: the spacetime metric $g_{\mu\nu}$, and a real scalar $\phi$ with its profile breaking conformal invariance that can be understood roughly as the running coupling of QCD. In addition to $\kappa_{N}^{2}$ that is the effective Newton constant, the potential $V(\phi)$ will be fixed by matching to the lattice QCD data.
	
The black hole with non-trivial scalar hair reads
 
\vspace{-5mm}
\begin{align}\label{ansatz}
\mathrm{d}s^2=-f(r) e^{-\eta(r)} \mathrm{d}t^2+\frac{\mathrm{d}r^2}{f(r)}+r^2\mathrm{d}\mathbf{x}_3^2,\quad
\phi=\phi(r)\,,
\end{align}
with $\mathrm{d}\mathbf{x}_3^2=\mathrm{d}x^2+\mathrm{d}y^2+\mathrm{d}z^2$ and $r$ the holographic radial coordinate. Denoting $r_h$ as the location of the event horizon where $f(r_h)=0$, the temperature reads $T=f'(r_h)e^{-\eta(r_h)/2}/4\pi$. Other thermodynamic quantities can be obtained straightforwardly using the standard holographic dictionary, see the Supplemental Material~\cite{SM} for more details. The next goal is to find a potential $V$ that can reproduce the EoS of $N_c=3$ pure gluon QCD. It comes as a nice surprise that the simple potential
	
\begin{figure*}
\centering
\includegraphics[width=0.44\textwidth]{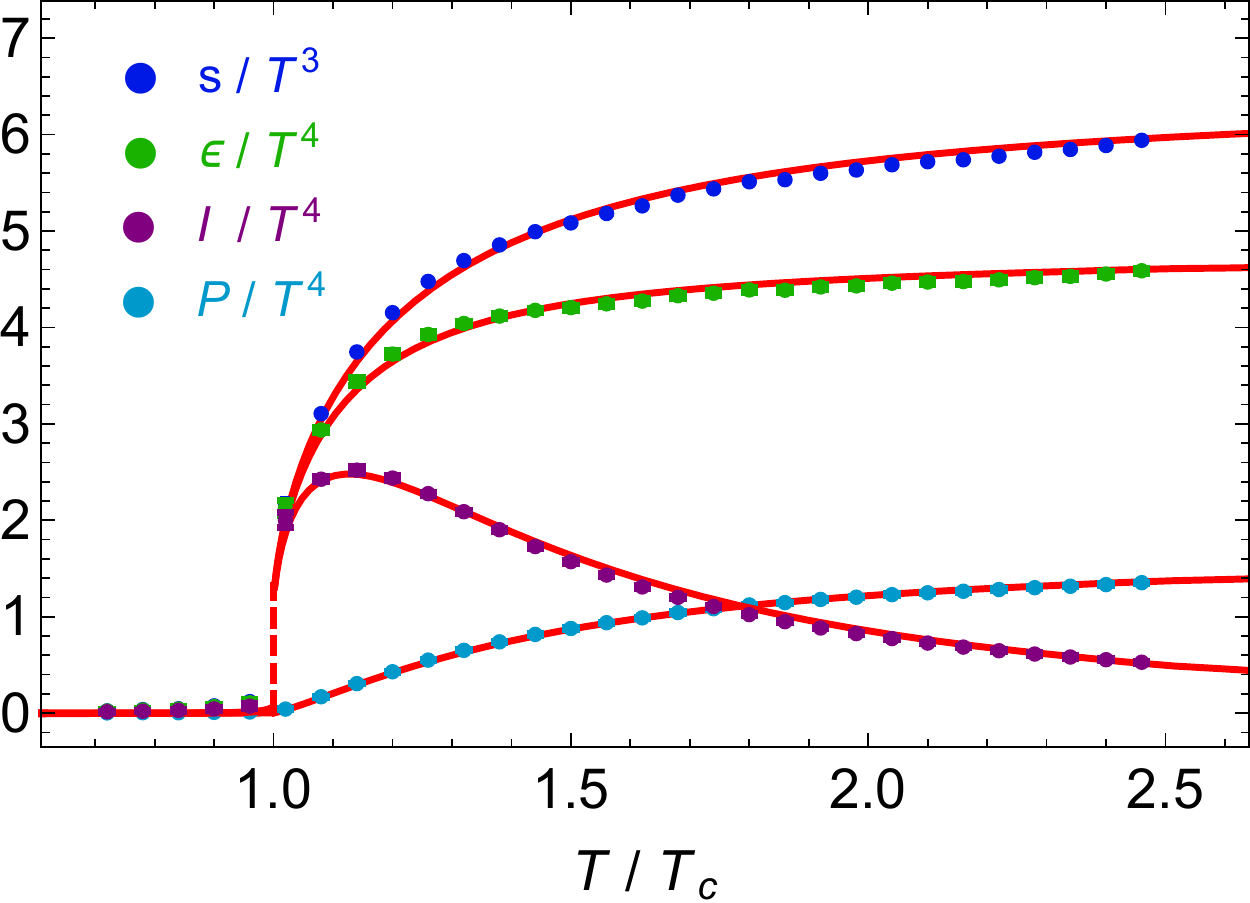}
\includegraphics[width=0.535\textwidth]{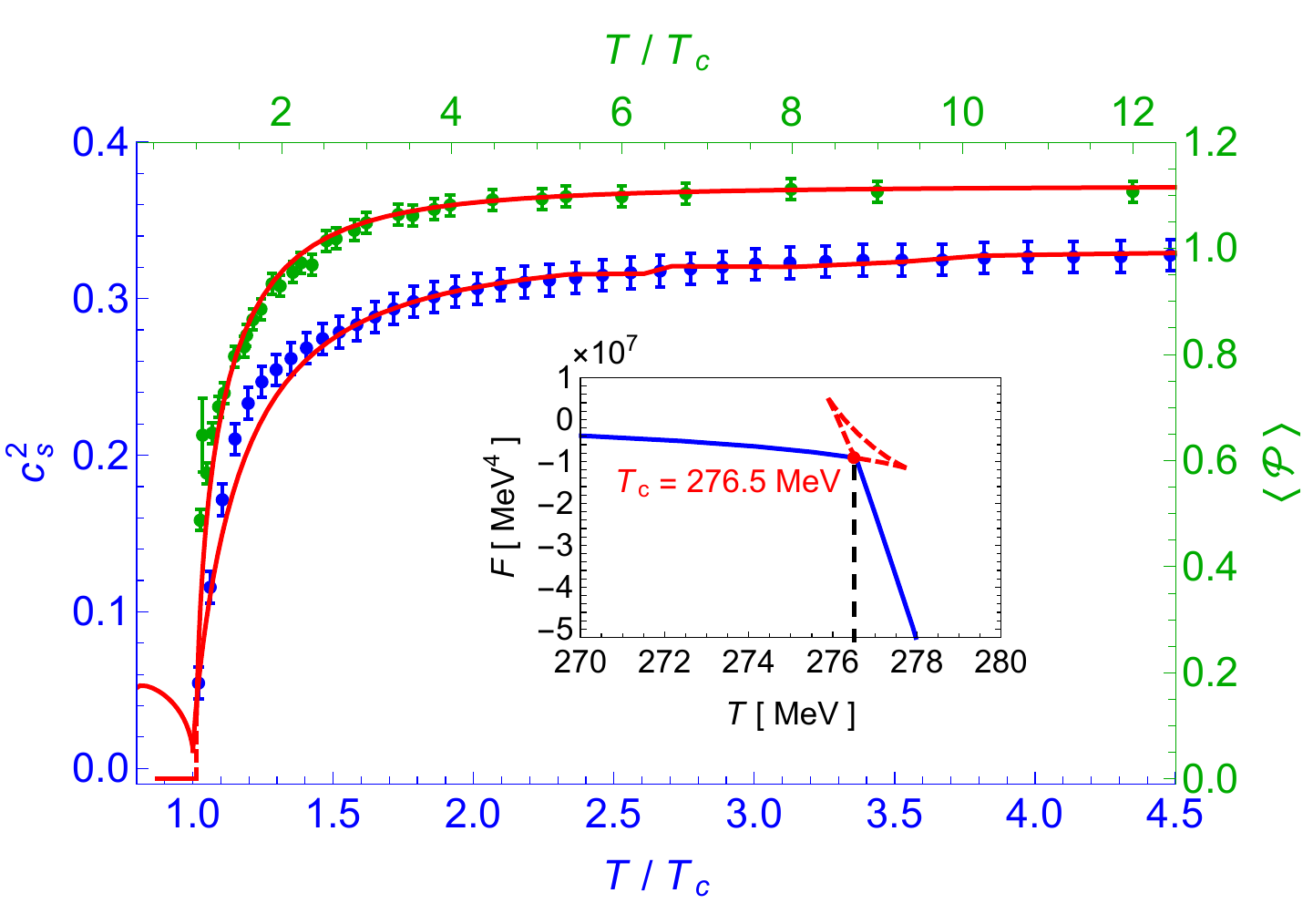}\\
\caption{The comparison between the lattice data (with error bar) of the pure gluon thermodynamics and our holographic calculations (solid curves) on various thermodynamic quantities. Discontinuous changes at the first-order phase transition are represented by dashed lines. \textbf{Left panel:} The temperature dependence of the energy density $\epsilon$, the entropy density $s$, the pressure $P$, and the trace anomaly $I=(\epsilon -3P)$~\cite{Caselle:2018kap}. \textbf{Right panel:} The squared speed of sound $c_s^2\equiv dP/d\epsilon$~\cite{Boyd:1996bx} and the Polyakov loop $\langle \mathcal{P} \rangle$~\cite{Gupta:2007ax} in function of temperature. \textbf{Insert:} The free energy density $F$ with respect to the temperature from our model. There is a first-order confinement/deconfinement phase transition at $T_c=276.5\,\text{MeV}$.}
\label{fig:lattice_comparison}
\end{figure*}

\vspace{-5mm}
\begin{align}\label{potential}
V(\phi)=\left(6 \gamma^2-\frac{3}{2}\right) \phi ^2-12 \cosh (\gamma \phi )
\end{align}
with $\gamma=0.735$ can reproduce the thermodynamics of lattice data for the pure gluon QCD \cite{Boyd:1996bx,Gupta:2007ax,Caselle:2018kap} as shown in Fig. \ref{fig:lattice_comparison}. Remarkably, although the error bars of the up-to-date lattice simulation \cite{Caselle:2018kap} are tiny, our theoretical results for EoS in the left panel are almost within these error bars. It is obvious from the free energy density $F$ that a strong FOPT takes place at the temperature $T_c=276.5$ MeV. We also compare the speed of sound $c_s$ in the right panel of Fig. \ref{fig:lattice_comparison}. Since $c_s$ is not provided in \cite{Caselle:2018kap}, we use the early data from lattice QCD \cite{Boyd:1996bx} and find good agreement.
	
\begin{figure}
\centering
\includegraphics[width=0.58\textwidth]{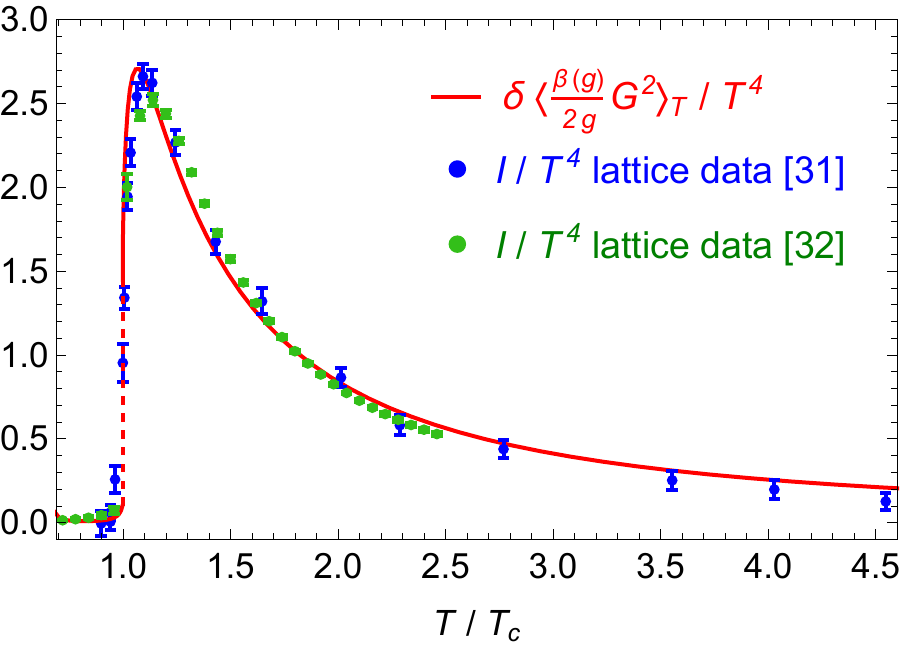}\\
\caption{The temperature dependence of the gluon condensation $\delta\left< \frac{\beta(g)}{2 g} G^2 \right>_T$ of our pure gluon model, where $\beta(g)$ is the $\beta$-function with $g$ the QCD gauge coupling. The data with error bar denotes the trace anomaly $I=(\epsilon-3 P)$ from lattice QCD~\cite{Boyd:1996bx,Caselle:2018kap}.}
\label{fig:glucond}
\end{figure}

 To understand the nature of the FOPT, we compute the expectation value of the Polyakov loop operator $\langle \mathcal{P} \rangle$ \cite{Polyakov:1997tj,Cai:2012xh,Colangelo:2010pe,Chen:2020ath}, which is a good order parameter to the deconfinement phase transition for pure gluon system~\cite{Fukushima:2017csk}. Surprisingly, $\langle \mathcal{P} \rangle$ by our holographic model quantitatively agrees with the lattice data~\cite{Gupta:2007ax} above $T_c$ and it quickly drops to zero below $T_c$, see the right panel of Fig.~\ref{fig:lattice_comparison}. It suggests that the FOPT from our model is a confinement/deconfinement phase transition. Remarkably, the temperature dependence of the gluon condensation $\delta\left< \frac{\beta(g)}{2 g} G^2 \right>_T$ capturing  the deconfinement phase transition is computed in our holographic model and is found to coincide with the trace anomaly $\epsilon-3p$ from EoS~\cite{Gubler:2018ctz,Cohen:1991nk}, see Fig.~\ref{fig:glucond}. Therefore, at $T_c$, we can then read off some essential quantities that are important to compute the SGWB and PBH productions associated with our FOPT.
The SGWBs generated in cosmological FOPTs were considered in other holographic models, see \emph{e.g.}~\cite{Cai:2022omk,He:2023ado,Ahmadvand:2017xrw,Bigazzi:2020avc,Li:2021qer}.
	
Independent of details of any specific particle physics model, the PBH production is a universal consequence of the FOPT~\cite{Liu:2021svg}. Due to the stochastic nature of bubble nucleations during FOPTs, the progress of populating true-vacuum bubbles in the false-vacuum background is {an asynochronized }process. There is always a non-vanishing probability to find some Hubble-sized regions to stay in the false vacuum for a slightly longer period of time than average. Since the radiation energy density should be rapidly diluted relative to the vacuum energy density in an expanding Universe, these Hubble-size regions would eventually accumulate enough overdensities in total energy density to finally reach the threshold of PBH productions. What is remarkable for this general mechanism of PBH productions during FOPTs is that the probability to find such Hubble-sized regions with postponed vacuum-decay progress can be made of particular observational interest for both detections from gravitational waves and PBHs, which will be briefly described shortly below and detailed in the Supplemental Material~\cite{SM}.
	
\section{Gravitational wave productions}
From the behavior of the free energy density in the insert of the right panel of Fig.~\ref{fig:lattice_comparison}, it clearly indicates the occurrence of a first-order confinement/deconfinement phase transition around the critical temperature $T_c=276.5$\, MeV, which could be a potentially important source for gravitational waves in the early Universe. The cosmological FOPT proceeds with stochastic nucleations of true vacuum bubbles in the false vacuum environment followed by the rapid expansion until percolations via bubble collisions. The bubble wall collision and plasma fluid motion including sound waves and magnetohydrodynamic (MHD) turbulences would generate the corresponding SGWBs with broken power-law shapes in their energy density spectra.
 
Given the expansion history $a(t)$ and vacuum decay rate of form $\Gamma(t)\equiv A(t)e^{-B(t)}$ per unit time and unit volume, the fraction of spatial regions that are still staying at the false vacuum at time $t$ can be estimated by~\cite{Guth:1982pn,Turner:1992tz}
 	
\vspace{-5mm}
\begin{align}
F(t;t_i)=\exp\left[-\frac{4\pi}{3}\int_{t_i}^t\mathrm{d}t'\,\Gamma(t')a(t')^3r(t,t')^3\right],
\end{align}
where $t_i$ is the earliest possible time for the nucleation of the first bubble ever, and $r(t,t')=\int_{t'}^t\mathrm{d}\tilde{t}/a(\tilde{t})$ is the comoving radius of a bubble at time $t$ nucleated from an earlier time $t'$. It is obvious that all regions are in the false vacuum before time $t_i$, namely $F(t<t_i; t_i)=1$. With the help of $F(t; t_i)$, the percolation time $t_*$ for the gravitational wave spectra from the FOPT is then conventionally defined by $F(t_*; t_i)=0.7$~\cite{percolation1971}, around which the decay rate can be expanded linearly in time for its exponent as $\Gamma(t)=A(t_*)e^{-B(t_*)+\beta(t-t_*)}\equiv\Gamma_0e^{\beta t}$~\cite{Cutting:2018tjt}\cite{footnote}. In this study, we will simply approximate the percolation temperature by the critical temperature $T_c=276.5$ MeV from previous holographic computations. See~\cite{Ares:2021nap} for a potential estimation on the effective potential and subsequent nucleation rate as well as the associated percolation temperature, which will be reserved for more detailed work in future.

The energy density spectra for the prementioned threes sources of SGWBs from a cosmological FOPT are given as follows. The uncollided part of bubble wall envelopes~\cite{Kosowsky:1991ua,Kosowsky:1992rz,Kosowsky:1992vn,Kamionkowski:1993fg,Huber:2008hg} admits an analytic form~\cite{Weir:2017wfa,Jinno:2016vai,Huber:2008hg} for the gravitational wave spectrum as
	
\vspace{-5mm}
\begin{align}
h^2\Omega_\mathrm{env}=1.67\times&10^{-5}\left(\frac{100}{g_\mathrm{dof}}\right)^\frac13\left(\frac{H_*}{\beta}\right)^2\left(\frac{\kappa_\phi\alpha}{1+\alpha}\right)^2
	\frac{0.48v_w^3}{1+5.3v_w^2+5v_w^4}S_\mathrm{env}(f),
\end{align}
where the shape factor is given by
 	
\vspace{-5mm}
\begin{align}
S_\mathrm{env}(f)=\frac{1}{c_l\left(\frac{f}{f_\mathrm{env}}\right)^{-3}+(1-c_l-c_h)\left(\frac{f}{f_\mathrm{env}}\right)^{-1}+c_h\left(\frac{f}{f_\mathrm{env}}\right)}
\end{align}
with $c_l=0.064$ and $c_h=0.48$, and the peak frequency is given by
 	
\vspace{-5mm}
\begin{align}
f_\mathrm{env}=1.65\times10^{-5}\,\mathrm{Hz}\left(\frac{g_\mathrm{dof}}{100}\right)^\frac16\left(\frac{T_*}{100\,\mathrm{GeV}}\right)\frac{0.35(\beta/H_*)}{1+0.069v_w+0.69v_w^4}.
\end{align}
Here the efficiency factor $\kappa_\phi$ characterizes the amount of released vacuum energy into the kinetic energy of bubble walls. The dominant contribution to the fluid motions comes from the sound waves~\cite{Hindmarsh:2013xza,Hindmarsh:2015qta,Hindmarsh:2017gnf}, whose gravitational wave spectrum is given by
 	
\vspace{-5mm}
\begin{align}
h^2\Omega_\mathrm{sw}=2.65\times&10^{-6}\left(\frac{100}{g_\mathrm{dof}}\right)^\frac13\left(\frac{H_*}{\beta}\right)\left(\frac{\kappa_\mathrm{sw}\alpha}{1+\alpha}\right)^2\frac{7^{7/2}v_w(f/f_\mathrm{sw})^3}{(4+3(f/f_\mathrm{sw})^2)^{7/2}}\Upsilon,
\end{align}
with the peak frequency
 	
\vspace{-5mm}
\begin{align}
f_\mathrm{sw}=1.9\times10^{-5}\,\mathrm{Hz}\left(\frac{g_\mathrm{dof}}{100}\right)^\frac16\left(\frac{T_*}{100\,\mathrm{GeV}}\right)\left(\frac{1}{v_w}\right)\left(\frac{\beta}{H_*}\right).
\end{align}
Here the efficiency factor $\kappa_\mathrm{sw}$ characterizes the amount of released vacuum energy into the kinetic energy of fluid motions, and the suppression factor $\Upsilon\equiv1-(1+2\tau_\mathrm{sw}H_*)^{-1/2}$~\cite{Guo:2020grp} accounts for the finite lifetime of sound waves from the onset timescale of turbulences, $\tau_\mathrm{sw}H_*\approx(8\pi)^{1/3}v_w/(\beta/H_*)/\bar{U}_f$ with the root-mean-squared fluid velocity given by $\bar{U}_f^2=3\kappa_\mathrm{sw}\alpha/[4(1+\alpha)]$. After the time scale $\tau_\mathrm{sw}$, the turbulences would eventually develop with gravitational wave spectrum analytically given by~\cite{Caprini:2009yp}
 	
	\vspace{-5mm}
	\begin{align}
	 h^2\Omega_\mathrm{tur}=3.35\times10^{-4}\left(\frac{100}{g_\mathrm{dof}}\right)^\frac13\left(\frac{H_*}{\beta}\right)\left(\frac{\kappa_\mathrm{tur}\alpha}{1+\alpha}\right)^\frac32 v_w S_\mathrm{tur}(f)\,,
	\end{align}
	where the efficiency factor $\kappa_\mathrm{tur}$ characterizes the amount of released vacuum energy into the turbulences, while the shape factor is given by
 	
	\vspace{-5mm}
	\begin{align}
	S_\mathrm{tur}(f)=\frac{(f/f_\mathrm{tur})^3}{[1+(f/f_\mathrm{tur})]^{\frac{11}{3}}(1+8\pi f/h_*)}
	\end{align}
	with the Hubble rate $h_*$ at $T_*$ given by
 	
	\vspace{-5mm}
	\begin{align}
	h_*=1.65\times10^{-5}\,\mathrm{Hz}\left(\frac{g_\mathrm{dof}}{100}\right)^\frac16\left(\frac{T_*}{100\,\mathrm{GeV}}\right),
	\end{align}
	and the peak frequency is given by
 	
	\vspace{-5mm}
	\begin{align}
	f_\mathrm{tur}=2.7\times10^{-5}\,\mathrm{Hz}\left(\frac{g_\mathrm{dof}}{100}\right)^\frac16\left(\frac{T_*}{100\,\mathrm{GeV}}\right)\left(\frac{1}{v_w}\right)\left(\frac{\beta}{H_*}\right).
	\end{align}
	\begin{figure*}
		\centering
		\includegraphics[width=0.48\textwidth]{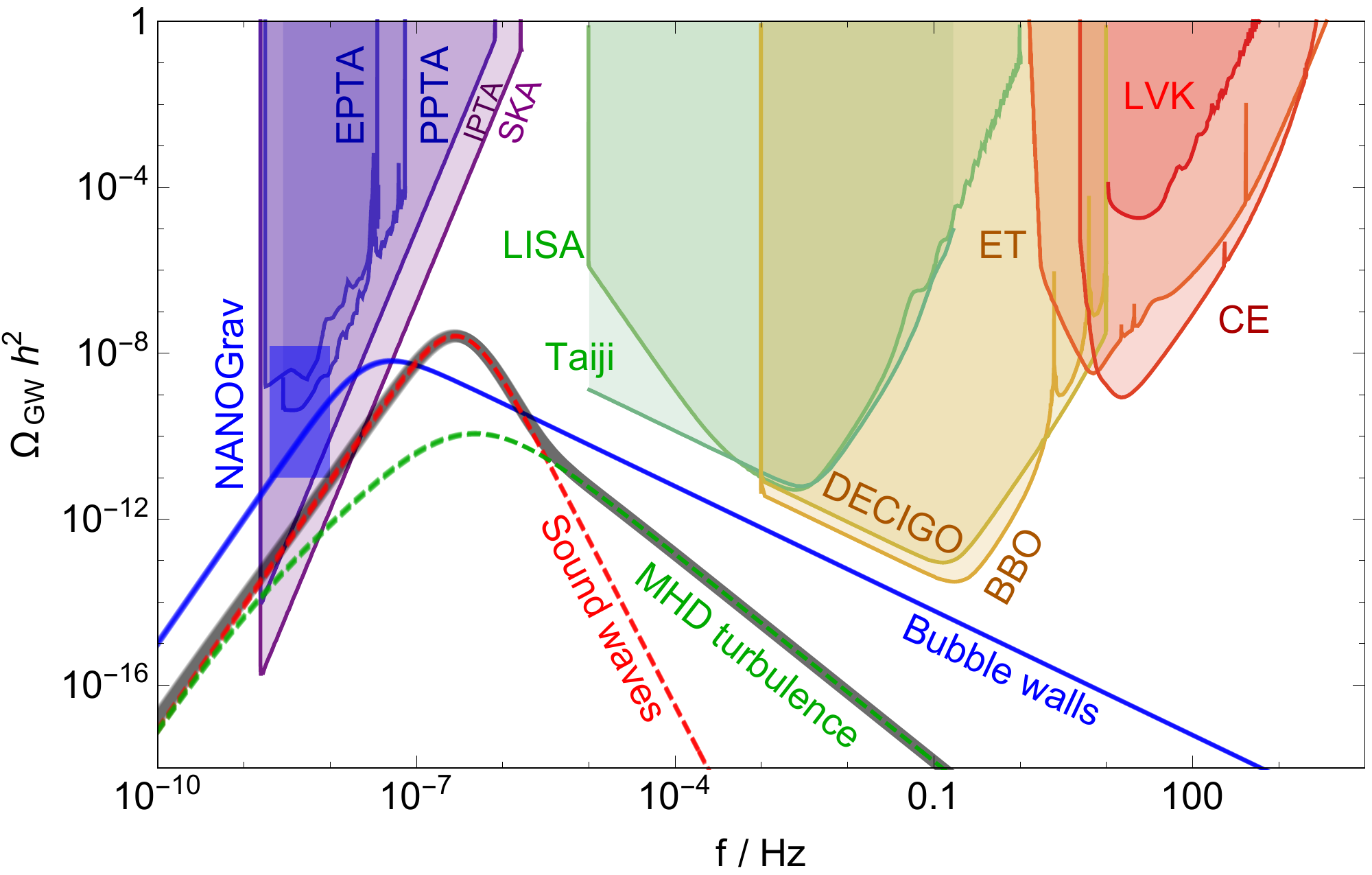}
		\includegraphics[width=0.48\textwidth]{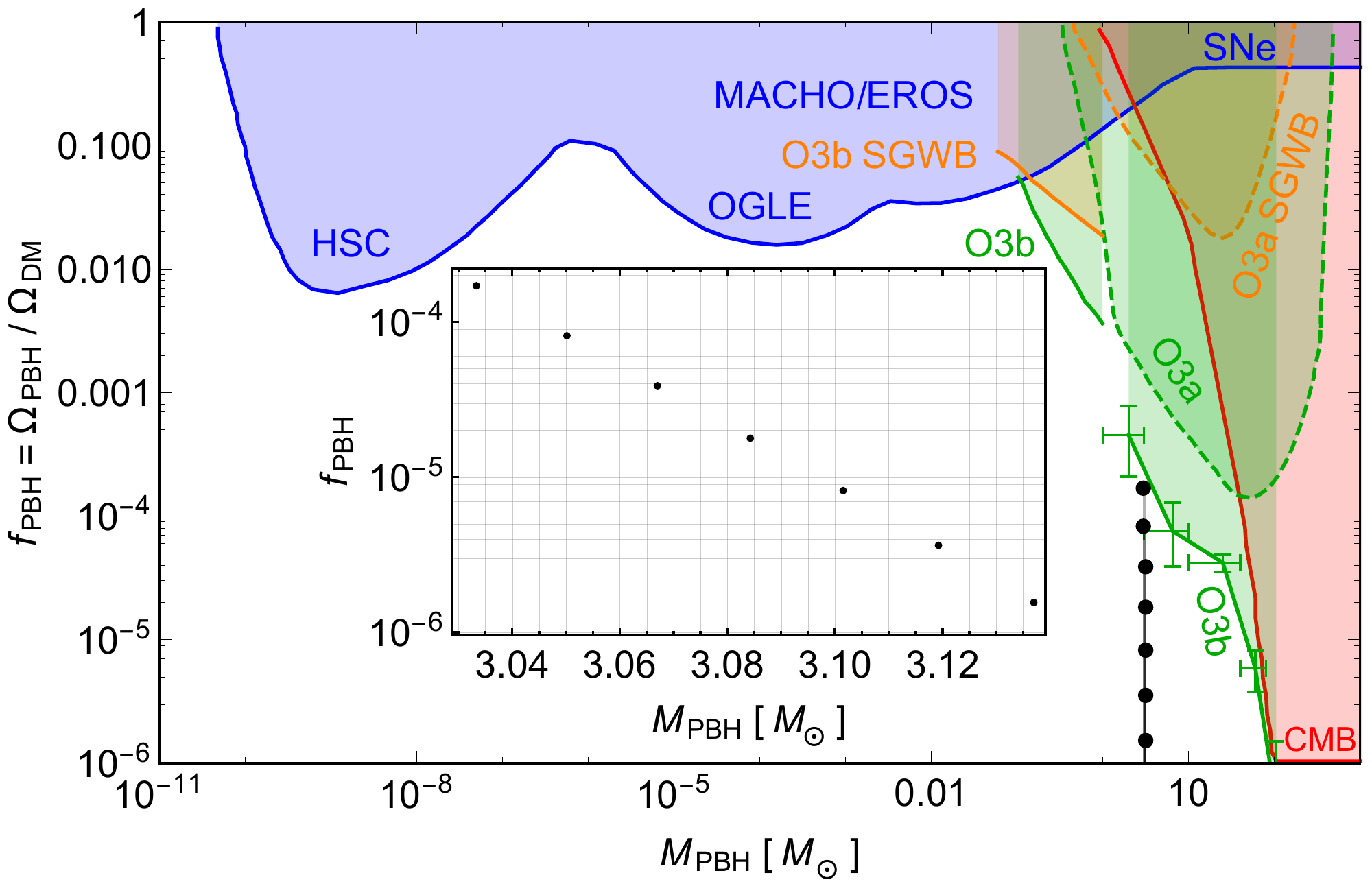}\\
		\caption{The predictions for SGWB (left) and PBH (right) productions from our holographic pure gluon model. In the left panel, the gravitational wave contributions from bubble walls (blue curve), sound waves (red curve), and MHD turbulences (green curve) are obtained for the associated largest PBH abundance allowed by the current observational constraints shown in the right panel. In the right panel, the current PBH constraints~\cite{Carr:2020xqk} are updated by including the constraints from LVC O3a data~\cite{Hutsi:2020sol} and O3b data~\cite{Nitz:2022ltl,Chen:2021nxo} (see~\cite{Franciolini:2022tfm} for a more recent constraint), and the insert zooms in the predicted PBH abundance regions. }
		\label{fig:PBHGWs}
	\end{figure*}

	For our holographic approach, the critical temperature $T_*=276.5$ MeV, the strength factor $\alpha\equiv\Delta V/\rho_r=0.939$, and the effective number of degrees of freedom $g_\mathrm{dof}\equiv\rho_r/\frac{\pi^2}{30}T_*^4=3.64$ are all fixed by the holographic thermodynamics, while the phase transition duration $\beta/H_*$, the bubble wall velocity $v_w$, and the efficiency factors $\kappa_\phi$, $\kappa_\mathrm{sw}$, and $\kappa_\mathrm{tur}$ are free parameters. To further fix some of the above parameters, we then consider two cases: (i) the envelope contribution would dominate the total gravitational wave spectrum for runaway walls (acceleration without termination at a constant velocity), however, runaway behavior has been recently argued~\cite{Bodeker:2009qy,Bodeker:2017cim,Hoche:2020ysm,Gouttenoire:2021kjv} to be improbable due to the growing friction force that eventually balances the driving force to reach a terminal wall velocity. Nevertheless, if the friction force is growing so slowly that most bubbles collide with each other before they even ever have a chance to reach the terminal velocity, the envelope contribution could still dominate the total gravitational wave spectrum~\cite{Cai:2020djd}. In this case, we can set the bubble wall velocity at collisions to be $v_w=1$ and the efficiency factors $\kappa_\phi=1$, $\kappa_\mathrm{sw}=0$, and $\kappa_\mathrm{tur}=0$ for a simple illustration. (ii) if most bubbles collide with each other at a constant terminal wall velocity, then the total gravitational wave spectrum is dominated by the fluid motions consisting of both sound waves and MHD turbulences. Hence, we can set the terminal wall velocity $v_w=0.95$ for a simple illustration. The efficiency factor $\kappa_\mathrm{sw}=\alpha/(0.73+0.083\sqrt{\alpha}+\alpha)$ for a relativistic wall velocity is originally obtained for a bag equation of state~\cite{Espinosa:2010hh}, but it changes insignificantly for a relativistic wall velocity and a large $\alpha$ even beyond a bag equation of state as shown in Fig.7 of Ref.~\cite{Wang:2022lyd} with a varying sound speed. We further set the efficiency factor $\kappa_\mathrm{tur}=0.1\kappa_\mathrm{sw}$ since some numerical simulations~\cite{Hindmarsh:2015qta,Caprini:2015zlo} seem to suggest that only at most $5\%-10\%$ of fluid motions is turbulent. While we are not able to compute $\beta/H_*$ from the first principle, it can be constrained by the PBH abundance associated with the FOPT. The SGWB spectra from our holographic model are shown in the left panel of Fig.~\ref{fig:PBHGWs} for case (i) (blue curve) and case (ii) (gray curve consisting of red and green dashed curves), where the expected sensitivity curves of future gravitational wave observatories are included. One can find that the SGWBs are within the reach of IPTA and SKA when the associated PBH abundance saturates the current observational bound from LVC constraints.
	
	\section{PBH productions}
	We then turn to the PBH productions. To evaluate the probability for the postponed vacuum decay, note that the differential probability for a Hubble-sized region $V_H(t)=\frac43\pi H(t)^{-3}$ not to decay at time $t$ reads $\mathrm{d}P(t)=1-\Gamma(t)V_H(t)\mathrm{d}t\approx \exp\left[-\Gamma(t)V_H(t)\mathrm{d}t\right]$, then the probability for this Hubble volume not to decay until time $t_n$ is obtained as
	
	\vspace{-5mm}
	\begin{align}
	P(t_n)=\prod\limits_{t=t_i}^{t_n}\mathrm{d}P(t)=\exp\left[-\int_{t_i}^{t_n}\mathrm{d}t\,V_H(t)\Gamma(t)\right]\,.
	\end{align}
	$P(t_n)$ is nothing but the PBH abundance $\Omega_\mathrm{PBH}$ at PBH formations if the overdensity in these Hubble volumes with postponed decay reaches the PBH formation threshold $\delta_c$,
	
	\vspace{-5mm}
	\begin{align}
	\delta(t_\mathrm{PBH})=\frac{\rho_r(t_\mathrm{PBH}; t_n)+\rho_v(t_\mathrm{PBH}; t_n)}{\rho_r(t_\mathrm{PBH}; t_i)+\rho_v(t_\mathrm{PBH}; t_i)}-1=\delta_c\,.
	\end{align}
	Here the vacuum energy densities inside and outside these Hubble volumes are estimated by $\rho_v(t;t_n)=F(t; t_n)\Delta V$ and $\rho_v(t;t_i)=F(t; t_i)\Delta V$, respectively, and the radiation energy densities inside and outside these Hubble volumes are solved from
	
	\vspace{-5mm}
	\begin{align}
	\frac{\mathrm{d}}{\mathrm{d}t}\rho_r(t;t_n)+4H(t;t_n)\rho_r(t;t_n)&=-\frac{\mathrm{d}}{\mathrm{d}t}\rho_v(t;t_n)\,,\\
	\frac{\mathrm{d}}{\mathrm{d}t}\rho_r(t;t_i)+4H(t;t_i)\rho_r(t;t_i)&=-\frac{\mathrm{d}}{\mathrm{d}t}\rho_v(t;t_i)\,,
	\end{align}
	respectively, where the Hubble parameters inside and outside these Hubble volumes are defined by $3M_\mathrm{Pl}^2H(t;t_n)^2=\rho_r(t;t_n)+\rho_v(t;t_n)$ and $3M_\mathrm{Pl}^2H(t;t_i)^2=\rho_r(t;t_i)+\rho_v(t;t_i)$, respectively. We adopt the analytic estimation~\cite{Harada:2013epa} on the PBH threshold $\delta_c=\sin^2[\pi\sqrt{w}/(1+3w)]=0.1786$ with the EoS $w\equiv P/\epsilon=0.0219$ evaluated from our holographic model for the dominant component at PBH formations. 
	
	Finally, the PBH abundance at matter-radiation equality is estimated by
	
	\vspace{-5mm}
	\begin{align}
	f_\mathrm{PBH}\equiv(a_\mathrm{eq}/a_\mathrm{PBH})\Omega_\mathrm{PBH}/\Omega_\mathrm{DM}(a_\mathrm{eq})\,,
	\end{align}
and the PBH mass is estimated as $M_\mathrm{PBH}=4\pi\gamma_\mathrm{PBH} M_\mathrm{Pl}^2/H_\mathrm{PBH}$ with usual PBH formation efficiency factor $\gamma_\mathrm{PBH}=0.2$~\cite{Carr:1975qj}. The PBH mass function $f_\mathrm{PBH}(M_\mathrm{PBH})$ is shown in the right panel of Fig.~\ref{fig:PBHGWs} with respect to current PBH constraints. It is worth noting that for PBH formations, apart from other phase transition parameters $T_*$, $\alpha$, and $g_\mathrm{dof}$ given by the holographic computations, the inverse of the phase transition duration $\beta$ is the only free parameter from our holographic approach, which could be constrained as $\beta/H_*>8.59$ from the current GWTC-3 data by $f_\mathrm{PBH}<0.00045$ in the mass range $[1\,M_\odot,3\,M_\odot]$~\cite{Chen:2021nxo}. Higher values for $\beta/H_*$ are certainly allowable like those in Refs.~\cite{Morgante:2022zvc,Chen:2022cgj} but with much more negligible PBH abundances and higher peak frequency and lower peak amplitude in the SGWBs that would be of less interest from both PBH and SGWB observations.
	
	\section{Conclusion and discussion}
	We have built up a holographic model for a pure gluon system to quantitatively confront the lattice data of $U(3)$ thermodynamics. It provides an effective model to capture the main feature of QCD matter, for which non-perturbative effects could be effectively adopted into the model parameters by matching with up-to-date lattice QCD (see also~\cite{Cai:2022omk,Li:2023mpv} for 2+1 flavors and~\cite{Zhao:2023gur} for 2 flavors). The resulting Polyakov loop operator and gluon condensation quantitatively match the lattice simulation, suggesting that there is a first-order confinement/deconfinement phase transition. The transition temperature is $T_c=276.5\text{MeV}$ as expected in lattice QCD literature. We have shown the gravitational wave energy spectrum and PBHs productions associated with the FOPT. With the most optimistic case constrained by the current PBH abundance, the energy spectrum of SGWBs could be potentially detectable within the sensitivity ranges of IPTA and SKA in the near future. Since finite volume effects in the lattice data have the tendency to smooth out the transition, our precise fitting to the lattice data might lead to an underestimate of the strength of the phase transition. Moreover, it is interesting to constrain our phase transition by the recently claimed common-spectrum red noise from PTA observations~\cite{Xu:2023wog,NANOGrav:2023gor,EPTA:2023sfo,Reardon:2023gzh} (see \emph{e.g.}~\cite{He:2023ado}).
	
	Since our holographic model can quantitatively capture the characteristic properties of the strong first-order confinement/deconfinement phase transition in a pure gluon system, one can study the transport properties in pure gluon and glueball gas to confirm the transition from a hydrodynamical point of view. It is worth considering real-time dynamics far from equilibrium, which is beyond the scope of lattice QCD. Moreover, it is an interesting direction to set up a holographic glueball action to compare the resulting glueball spectra with more experimental and lattice data. By appropriately tuning the scalar potential $V(\phi)$ to match the lattice data offered by \cite{Panero:2009tv}, our setup could in principle be applied to model a strongly coupled $SU(N)$ dark sector (see \cite{Guo:2022puy} for a recent study on hot $SU(N)$). Lacking a top-down string embedding, the precisely meaning of the operator dual to the bulk scalar $\phi$ is not clear. Nevertheless, our findings for the pure gluon in the present work as well as the (2+1)-flavor~\cite{Cai:2022omk,Li:2023mpv} and 2-flavor ~\cite{Zhao:2023gur} cases suggest that the dual operator with dimension 3 could play an important role in describing QCD dynamics in the non-perturbative regime. It would be interesting to find out this operator in benchmark effective theories and low-energy models of QCD. Furthermore, the present results,  particularly those regarding the confinement/deconfinement phase transition, should be embedded into the framework of a general and hybrid QCD phase diagram, including, \emph{e.g.}, an external magnetic field and a rotation. 
	
	\section*{Acknowledgments}
	We thank Qing-Guo Huang, Yi-Fan Wang, Zu-Cheng Chen, Shi Pu, Yong Cai and Peng Liu for stimulating discussions. This work was supported by National Key Research and Development Program of China Grant No. 2021YFC2203004 and No. 2020YFC2201501, the National Natural Science Foundation of China Grants No. 12075101, No. 12235016, No. 12122513, No. 12075298, No. 12047569, No. 11991052, No. 12047503, No. 12105344, No. 11947233, and No. 12235019, the Key Research Program of the Chinese Academy of Sciences (CAS) Grant NO. XDPB15, and the Science Research Grants from the China Manned Space Project with No. CMS-CSST-2021-B01.
	S.H. also would like to appreciate the financial support from Jilin University and Max Planck Partner group. We also acknowledge the use of the HPC Cluster of ITP-CAS.


\bibliographystyle{utphys}
\bibliography{ref}

\begin{thebibliography}{10}
\bibitem{footnote}
Note that the exponent in the decay rate
  $\Gamma(t)=A(t)e^{-B(t)}=\exp[-B(t)+\ln A(t)]$ can be expanded around $t_*$
  to the first order, $-B(t_*)+\ln A(t_*)+\beta(t-t_*)+(\mathrm{d}\ln
  A(t_*)/\mathrm{d}t)(t-t_*)$, where the time dependence in the last derivative
  term is suppressed by the logarithmic function, and hence we can approximate
  $\Gamma(t)\approx[A(t_*)e^{-B(t_*)-\beta t_*}]e^{\beta t}\equiv\Gamma_0
  e^{\beta t}$.
\end{thebibliography}

\begin{appendix}

\section{Thermodynamics and model parameters fixing}

We start from the following 5-dimensional gravitational action.
\begin{align}\label{action1}
S=\frac{1}{2\kappa_N^2}\int \mathrm{d}^{5}x \sqrt{-g} \left[\mathcal{R}-\frac{1}{2}\nabla_\mu \phi \nabla^\mu \phi-V(\phi)\right]\,.
\end{align}
The gravitational theory includes only two fields: the spacetime metric $g_{\mu\nu}$, and a real scalar $\phi$. In addition, $\kappa_{N}^{2}$ is the effective Newton constant and the $V(\phi)$ is the scalar potential.

The black hole with non-trivial scalar hair reads
\begin{align}\label{ansatz}
\mathrm{d}s^2=-f(r) e^{-\eta(r)} \mathrm{d}t^2+\frac{\mathrm{d}r^2}{f(r)}+r^2\mathrm{d}\mathbf{x}_3^2,\quad
\phi=\phi(r)\,,
\end{align}
with $\mathrm{d}\mathbf{x}_3^2=\mathrm{d}x^2+\mathrm{d}y^2+\mathrm{d}z^2$ and $r$ the holographic radial coordinate. Substituting the ansatz~\eqref{ansatz}, we obtain the following independent equations of motion (EoM).
\begin{align}\label{eoms}
\begin{split}
\phi''+\left(\frac{f'}{f}-\frac{\eta'}{2}+\frac{3}{r}\right)\phi'-\frac{1}{f}\partial_\phi V& = 0\,, \\
\frac{\eta'}{r}+\frac{1}{3}\phi'^2&=0\,,\\
\frac{2}{r}\frac{ f'}{f}-\frac{\eta'}{r}+\frac{2}{3 f}V+\frac{4}{r^2}&=0\,,
\end{split}
\end{align}
where the prime denotes the derivative with respect to $r$. In what follows we will specify $V(\phi)$ as
\begin{align}\label{app:potential}
V(\phi)=\left(6 \gamma^2-\frac{3}{2}\right) \phi ^2-12 \cosh (\gamma \phi )\,,
\end{align}
where $\gamma$ is the only free parameter. Note, however, that to fit the EoS for (2+1)-flavor QCD at zero baryon density, one has to introduce two free parameters~\cite{Cai:2022omk} and three free parameters~\cite{Grefa:2021qvt} in $V(\phi)$.

Near the AdS boundary $r\rightarrow\infty$ where $\phi\rightarrow 0$, one has
\begin{align}
V(\phi)=-12-\frac{3}{2}\phi^2+\mathcal{O}(\phi^4)\,.
\end{align}
Therefore, the cosmological constant is given by $\Lambda=-6$ (the AdS radius $L=1$) and the scaling dimension of the dual scalar operator is $\Delta=3$. We then obtain the following asymptotic expansion:
\begin{align}\label{uvexpand}
\begin{split}
\phi(r)& = \frac{\phi _s}{r}+\frac{\left(\gamma^4-1/6\right) \phi _s^3\ln r+\phi _v}{r^2}+...\,,\\
\eta(r)&=\frac{\phi _s^2}{6r^2}+\frac{(1-6 \gamma^4) (1-12\ln r)\phi _s^4+72 \phi _s \phi _v}{144 r^4}+...,\\
f(r)&=r^2+\frac{\phi _s^2}{6}+\frac{2 f_v-\phi _s^4(1-6\gamma^4)\ln r}{12 r^2}+...\,,
\end{split}
\end{align}
where we have taken the normalization such that $\eta(r\rightarrow\infty)=0$. $\phi_s$ is the source of the scalar operator of the boundary theory, which essentially breaks the conformal symmetry and plays the role of the energy scale.

To read off the physical observables, we incorporate the holographic renormalization by adding the boundary terms that are given as~\cite{Li:2020spf}
\begin{align}\label{byterm}
S_\partial =\frac{1}{2\kappa_N^2}\int\mathrm{d}x^4\left[2K-6-\frac{\phi^2}{2}-\left(b+\frac{6\gamma^4-1}{12} \ln r\right)\phi^4\right]
\end{align}
at the AdS boundary $r\rightarrow\infty$. Here $h_{\mu\nu}$ is the induced metric and $K_{\mu\nu}$ is the extrinsic curvature defined by the outward pointing normal vector to the boundary.

The energy-momentum tensor of the dual boundary theory reads
\begin{align}\label{Tmunu}
\begin{split}
T_{\mu\nu}&=\lim_{r\rightarrow\infty}\frac{2\,r^2}{\sqrt{-h}} \frac{\delta (S+S_\partial)_{on-shell}}{\delta h^{\mu \nu}}\\
&=\frac{1}{2\kappa_N^2}\lim_{r\rightarrow\infty}r^2\left[2(K h_{\mu\nu}-K_{\mu\nu}-3 h_{\mu\nu})
-\left(\frac{1}{2}\phi^2+\frac{6c_1^4-1}{12}\phi^4 \ln r+b\,\phi^4\right)h_{\mu\nu}\right].
\end{split}
\end{align}
Inserting the UV expansion~\eqref{uvexpand}, we obtain
\begin{align}\label{EP}
\begin{split}
\epsilon &\equiv T_{tt}=\frac{1}{2\kappa_N^2}\left(-3 f_v+\phi_s\phi_v+\frac{1+48 b}{48}\phi_s^4\right),\\
P&\equiv T_{xx}=\frac{1}{2\kappa_N^2}\left(-f_v+\phi_s\phi_v+\frac{3-48 b-8\gamma^4}{48}\phi_s^4\right),\\
I&\equiv\epsilon-3P=\frac{1}{2\kappa_N^2}\left(-2\phi_s\phi_v-\frac{1-24 b-3\gamma^4}{6}\phi_s^4\right).
\end{split}
\end{align}
The temperature and entropy density are given by
\begin{align}
T=\frac{1}{4\pi}f'(r_h)e^{-\eta(r_h)/2},\quad s=\frac{2\pi}{\kappa_N^2} r_h^3\,,
\end{align}
where $r_h$ is the location of the event horizon.

The free energy density $F$ is identified as the temperature $T$ times the renormalized action in the Euclidean signature.
\begin{equation}\label{freeE}
F=\frac{T}{V} (S+S_\partial)_\mathrm{on-shell}^{\text{Euclidean}}=\frac{1}{2\kappa_N^2}\left(f_v-\phi_s\phi_v-\frac{3-48 b-8\gamma^4}{48}\phi_s^4\right)\,.
\end{equation}
with $V=\int\mathrm{d}x\mathrm{d}y\mathrm{d}z$. Taking advantage of radially conserved quantity
\begin{align}
\mathcal{Q}=\frac{1}{2\kappa_N^2}r^5 e^{\eta/2}\left(\frac{f}{r^2}e^{-\eta}\right)',
\end{align}
and then evaluating at both horizon $r=r_h$ and UV boundary $r\rightarrow\infty$, we obtain the expected thermodynamic relation
\begin{align}
F=\epsilon-T\,s=-P\,.
\end{align}

After obtaining the thermodynamic quantities, one can also compute some important transport coefficients, such as the speed of sound $c_s=\sqrt{\mathrm{d}P/\mathrm{d}\epsilon}$. These quantities are compared to the lattice results for pure gluon~\cite{Boyd:1996bx,Gupta:2007ax,Caselle:2018kap}. Then all free parameters of our holographic model can be fixed to be
\begin{equation*}
\gamma=0.735,\; \kappa_{N}^2 =9.76 \pi,\; \phi_s=1.523 \text{GeV},\; b=0.06777\,.
\end{equation*}
The last parameter $b$ that appears in the boundary term~\eqref{byterm} corresponds to $P(T=0)=0$. The fitting results are presented in Fig.~[1] in the main text, from which there is a first-order phase transition (FOPT) at $T_c=276.5\,\text{MeV}$.

\section{Computations of gluon condensation}

To study the gluon condensation in our pure gluon model, we adopt a probe scalar field $\chi(r)$ on the background~\eqref{ansatz}. The action reads
\begin{equation}
S=\frac{1}{2 \kappa_N^2}\int \mathrm{d}^5 x \sqrt{-g^s} e^{-\sqrt{\frac{3}{8}}\phi}\left[-\frac{1}{2}\nabla_\mu \chi \nabla^\mu \chi -\frac{1}{2}m_{\chi}^2 \chi^2\right].
\end{equation}
Here $g^s$ is the determinant of the metric in the string frame with
\begin{equation}
g^s_{\mu \nu}=e^{\sqrt{\frac{2}{3}}\phi}g_{\mu \nu},
\end{equation}
where $g_{\mu \nu}$ is the metric in the Einstein frame used in the previous section.

Then, the EoM of $\chi(r)$ is
given by
\begin{equation}
\chi''+\frac{1}{4} \left( \frac{12}{r}+\frac{4 f'}{f}-2\eta'+\sqrt{6}\phi' \right)\chi'-\frac{e^{\sqrt{\frac{2}{3}}\phi}}{f}m_{\chi}^2\chi=0\,.
\end{equation}
One considers the regular boundary condition on the IR:
\begin{equation}
\chi(r)=c_0+c_1(r-r_h)+c_2(r-r_h)^2+\dots\,,
\end{equation}
The UV expansion shows
\begin{equation}
\chi(r)=\chi_{0} r^{\Delta-4}+\dots+\chi_{4} r^{-\Delta}+\dots\,.
\end{equation}
The source $\chi_0$ will be fixed to be a constant so as to fit the lattice data. The holographic renormalized gluon condensation reads
\begin{equation}
\chi_4=\left<G^2\right>\,.
\end{equation}
On the other hand, the subtracted gluon condensation $\delta\left< \frac{\beta(g)}{2 g} G^2 \right>_{T}$ is related to the trace anomaly~\cite{Gubler:2018ctz,Cohen:1991nk}
\begin{equation}\label{gluc}
\delta\left< \frac{\beta(g)}{2 g} G^2 \right>_{T}\equiv\left< \frac{\beta(g)}{2 g} G^2 \right>_{T}-\left< \frac{\beta(g)}{2 g} G^2 \right>_{0}=\epsilon-3 P\,,
\end{equation}
where the coefficient $\beta(g)$ is the $\beta$-function of QCD and $\left< \frac{\beta(g)}{2 g} G^2 \right>_{0}$ is determined by the exploration value of $\left< \frac{\beta(g)}{2 g} G^2 \right>_{T}$ from finite temperature. For simplicity, we directly call $\delta\left< \frac{\beta(g)}{2 g} G^2 \right>_{T}$ as physical gluon condensation in the main text. Following~\cite{Gubser:2008yx}, we choose the renormalized dimension of the gluon operator $\Delta=3.93$, which in turn determines the mass of bulk scalar via $m_{\chi}^2=\Delta(\Delta-4)$.

The $\beta$-function to 3-loop reads
\begin{equation}
\beta (g)=-\beta_0 g^3-\beta_1 g^5-\beta_2 g^7+O(g^9)\,.
\end{equation}
Therefore, the coefficient in~\eqref{gluc} are given by
\begin{equation}
\frac{ \beta(g)}{2 g}=-\left(2 \pi \beta_0 \alpha_s+8 \pi^2 \beta_1 \alpha_s^2+32\pi^3 \beta_2 \alpha_s^3\right)\,,
\end{equation}
with~\cite{Alekseev:2002zn}
\begin{align}
\alpha_s(T)=\frac{1}{4\pi \beta_0 Q}\left[ 1-\frac{\beta_1}{\beta_0^2}\frac{\mathrm{ln}Q}{Q}+\frac{\beta_1^2}{\beta_0^4 Q^2}\left(\left(\mathrm{ln}Q\right)^2-\mathrm{ln}Q-1+\frac{\beta_0\beta_2}{\beta_1^2}\right) \right]\,,
\end{align}
where $Q=\mathrm{ln}(T^2/\Lambda_{QCD}^2)$ and
\begin{align}
&\beta_0=\frac{1}{(4\pi)^2}\left(11-\frac{2}{3}N_f\right),\qquad \beta_1=\frac{1}{(4\pi)^4}\left(102-\frac{38}{3}N_f\right),\,\nonumber\\
& \beta_2=\frac{1}{\left(4\pi\right)^6}\left(\frac{2857}{2}-\frac{5033}{18}N_f+\frac{325}{54}N_f^2\right).
\end{align}
The cutoff $\Lambda_{QCD}$ gives the effective range of energy scale for $\alpha_s(T)$ which means $\alpha_s(T)$ only works for $T>\Lambda_{QCD}$.
The temperature dependent gluon condensation is shown in Fig.[2] with $\Lambda_{QCD}=0.14\mathrm{GeV}$ and $\chi_0=-2.97\times 10^{-10}\mathrm{GeV}^{4-\Delta}$. The data for the trace anomaly from lattice QCD~\cite{Boyd:1996bx,Caselle:2018kap} is included. It is clearly that our results match the lattice data pretty well.

\section{Polyakov Loop}
With the gravitational background~\eqref{ansatz}, one can extract the expectation value of the Polyakov loop operator~\cite{Polyakov:1997tj} in terms of the holographic dictionary. For simplicity, one makes the coordinate transformation $z=\frac{1}{r}$ and field redefinition $f(r)=\frac{F(z)}{z^2}, \eta(r)=\Sigma(z), \phi(r)= \Phi(z)$. The background~\eqref{ansatz} now becomes
\begin{equation}
\mathrm{d}s^2=\frac{1}{z^2}\left[-F(z) e^{-\Sigma(z)} \mathrm{d}t^2+\frac{\mathrm{d}z^2}{F(z)}+d\boldsymbol{x_3^2}\right].\label{eq:metric}
\end{equation}
The world-sheet action for the Polyakov loop~\cite{Li:2011hp,Cai:2012xh} in the string frame reads
\begin{equation}\label{NG}
\begin{split}
S_{NG}=\frac{1}{2\pi \alpha_p}\int \mathrm{d}^2\xi \,  e^{\sqrt{\frac{2}{3}}\Phi(z)}\sqrt{\mathrm{det}[g_{MN}(\partial_a X^M)(\partial_b X^N)]}\,,
\end{split}
\end{equation}
where $g_{MN}$ is the target spacetime metric and $g_{ab}$ is the induced metric. Here $\alpha'$ is the effective string tension.

Without loss of generality, we choose the static gauge condition $\xi^0=t$, $\xi^1=x$, $z=z(x)$, from which
\begin{equation}
g_{ab}=
\left(
\begin{array}{cc}
-\frac{ F(z) e^{-\Sigma(z)}}{z^2} & 0\\
0 & \frac{1}{z^2}\left(\frac{z'(x)^2}{F(z)}+1\right)
\end{array}
\right)\,.
\label{eq:induced metric}
\end{equation}
Thus, the action~\eqref{NG} becomes
\begin{equation}
\begin{split}
S_{NG}&=\frac{1}{2\pi \alpha_p}\int \mathrm{d}t \int_{-\frac{\bar{r}}{2}}^{\frac{\bar{r}}{2}} \, \mathrm{d}x \frac{e^{\sqrt{\frac{2}{3}}\Phi(z)-\frac{\Sigma(z)}{2}}}{z^2}\sqrt{F(z)+z'(x)^2}\,,\\
&=\frac{{1}}{\pi \alpha_p T} \int_{-\frac{\bar{r}}{2}}^{0} \, \mathrm{d}x \frac{e^{\sqrt{\frac{2}{3}}\Phi(z)-\frac{\Sigma(z)}{2}}}{z^2}\sqrt{F(z)+z'(x)^2} \,,
\end{split}
\label{eq:NG}
\end{equation}
with the boundary conditions
\begin{equation}
z(x=0)=z_0,\quad z'(x=0)=0,\quad z(x=\pm\frac{\ell}{2})=0\,.
\end{equation}
Here $\ell$ is the endpoint of the string on the boundary.
The equation of motion is given by
\begin{equation}
\begin{split}
z'(x)^2=F(z) \left(e^{-\Sigma(z)+\Sigma(z_0)+2\sqrt{\frac{2}{3}}(\Phi(z)-\Phi(z_0))}(\frac{z_0}{z})^4\frac{F(z)}{F(z_0)}-1\right).
\end{split}
\end{equation}
Plugging it into~\eqref{eq:NG} and introducing the coordinate transformation $v=\frac{z}{z_0}$, we obtain
\begin{equation}
\begin{split}
S_{NG}&=\frac{{1}}{\pi \alpha_p T}\int_{0}^{z_0} \, \mathrm{d}z \frac{e^{\sqrt{\frac{2}{3}}\Phi(z)-\frac{\Sigma(z)}{2}}}{z'(x) z^2}\sqrt{F(z)+z'(x)^2}\,,\\
&=\frac{{1}}{\pi \alpha_p z_0 T}\int_{0}^{1} \, \mathrm{d}v \frac{e^{-\frac{1}{2}\Sigma(v z_0)+\sqrt{\frac{2}{3}} \Phi(v z_0)}}{v^2 \, \tau(v)}\,,
\end{split}
\label{eq:NGfinal}
\end{equation}
where $\tau(v)$ is
\begin{equation}
\begin{split}
\tau(v)=\sqrt{1-\frac{e^{-\Sigma(z_0)+\Sigma(v z_0)-2\sqrt{\frac{2}{3}}(-\Phi(z_0)+  \Phi(v z_0)) }v^4 F(z_0)}{F(v z_0)}}\,.
\end{split}
\end{equation}
The on-shell renormalized free energy for the Polyakov loop operator is given by
\begin{equation}\label{Polyakovloop}
F_p=\frac{1}{\pi \alpha_p z_0}\left[-1+\int_{0}^{1} \, \frac{\mathrm{d}v}{v^2} \left(\frac{e^{-\frac{1}{2}\Sigma(v z_0)+\sqrt{\frac{2}{3}} \Phi(v z_0)}}{\tau(v)}-1\right)\right]\,.
\end{equation}
Then the expectation value of the Polyakov loop operator shows \cite{Cai:2012xh}
\begin{equation}\label{EVPolyakovloop}
\langle \mathcal{P} \rangle=e^{C_p-\frac{F^{\infty}_p}{2T}}
\end{equation}
with $F^{\infty}_p$ the maximum of $F_p$ \cite{Colangelo:2010pe} and $C_p$ the normalization constant~\cite{Bak:2007fk,Li:2011hp}. 

The expectation value of Polyakov loop operator $\langle \mathcal{P} \rangle$ is a good order parameter to the deconfinement PT for pure gluon system~\cite{Fukushima:2017csk}. The computation of Polyakov loops in holography was given in~\cite{Bak:2007fk}. Note that here we adopt the effective string tension $\alpha_p=17.5$ and the renormalization constant $C_p=0.11$~\cite{Li:2011hp}.
Surprisingly, as shown in the right panel of Fig.~1 in the main text,  the temperature dependence of $\langle \mathcal{P} \rangle$ from our model quantitatively agrees with the lattice data~\cite{Gupta:2007ax} above $T_c$ and drops to zero below $T_c$, suggesting that the FOPT from our model is a confinement/deconfinement phase transition (PT).


\section{Computations of GWs and PBHs}

\begin{figure}
	\centering
	\includegraphics[width=0.45\textwidth]{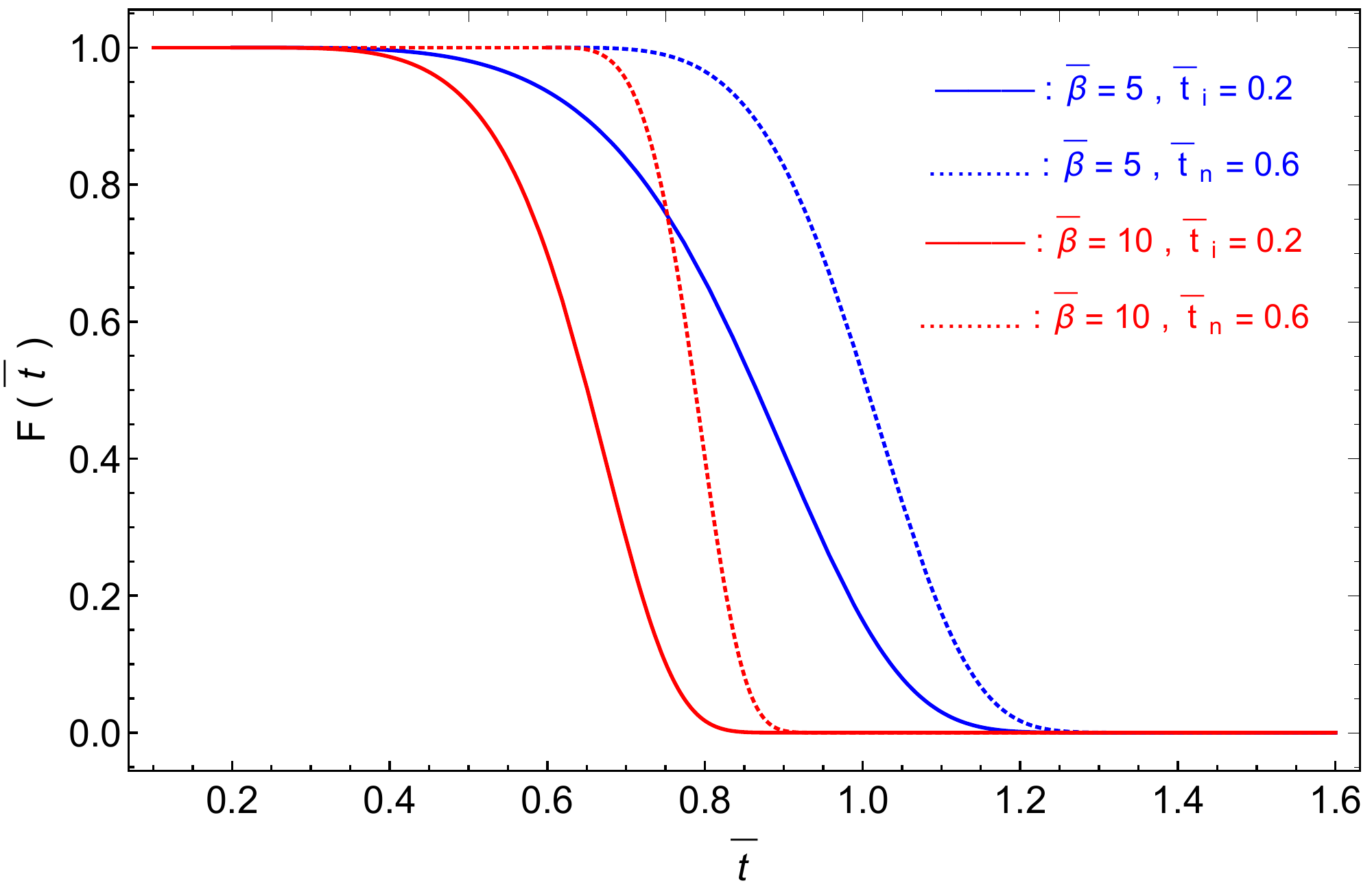}\\
	\includegraphics[width=0.45\textwidth]{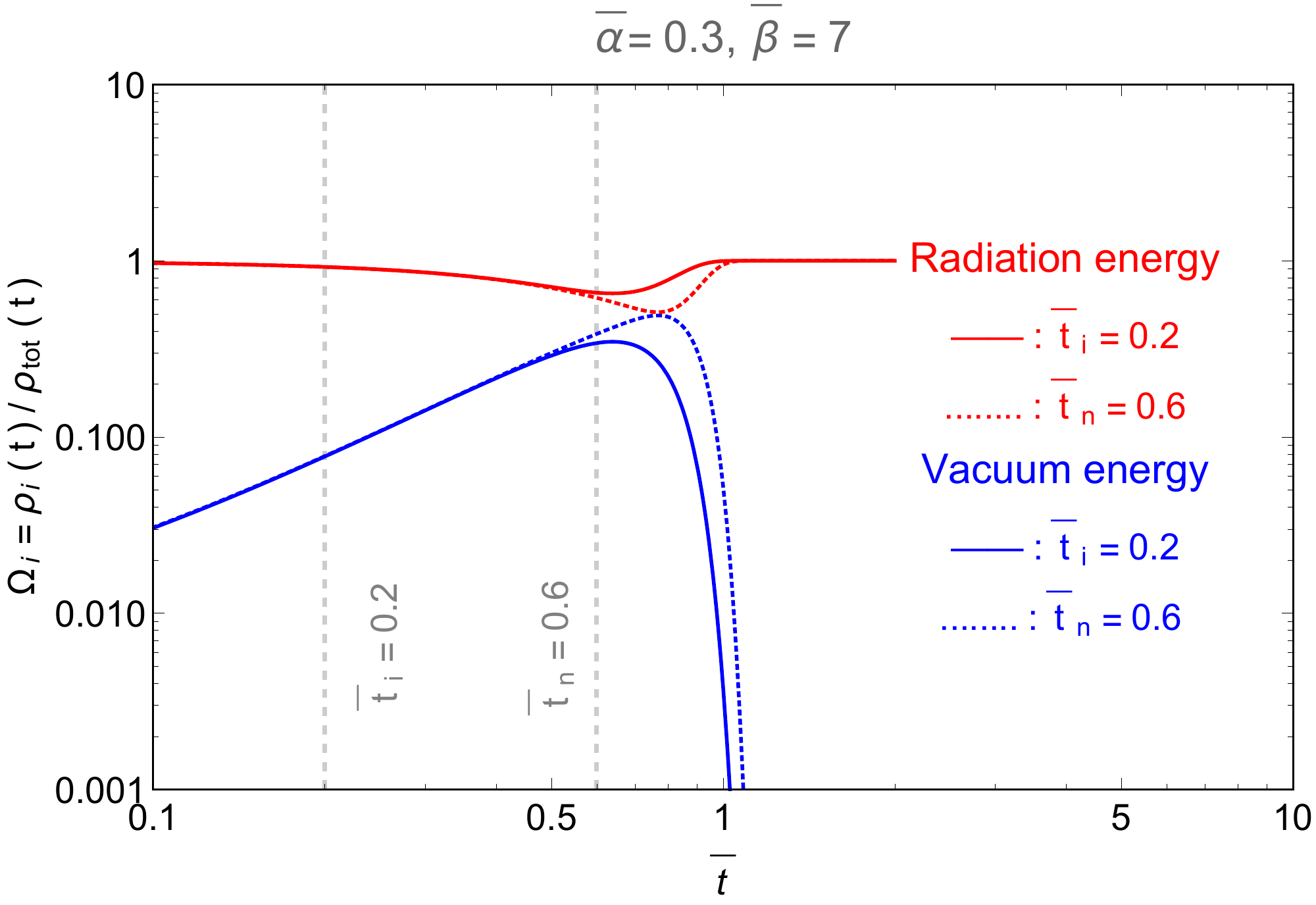}
	\includegraphics[width=0.45\textwidth]{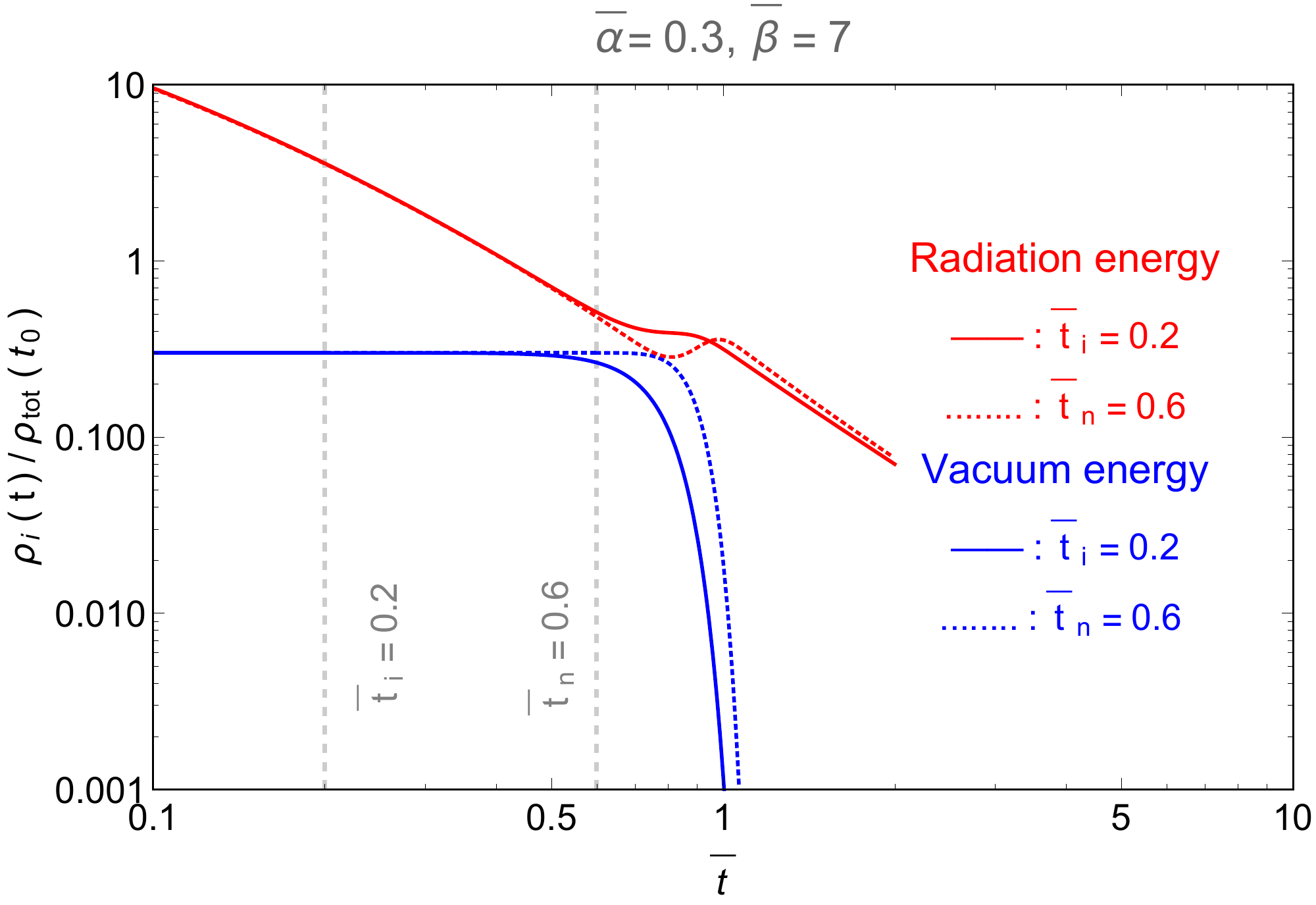}\\
	\caption{The time evolution for the spatial fraction of false vacuum regions (top), the radiation and vacuum energy density  fractions (bottom left) and the radiation and vacuum energy densities normalized by the total energy density at some time $\bar{t}_0=1/2$ (bottom right). In all panels, $\bar{t}_i=0.2$ and $\bar{t}_n=0.6$ denote the normal decay channel and delayed decay channel, respectively. }
	\label{fig:TimeEvolution}
\end{figure}

\begin{figure*}
	\centering
	\includegraphics[width=0.4\textwidth]{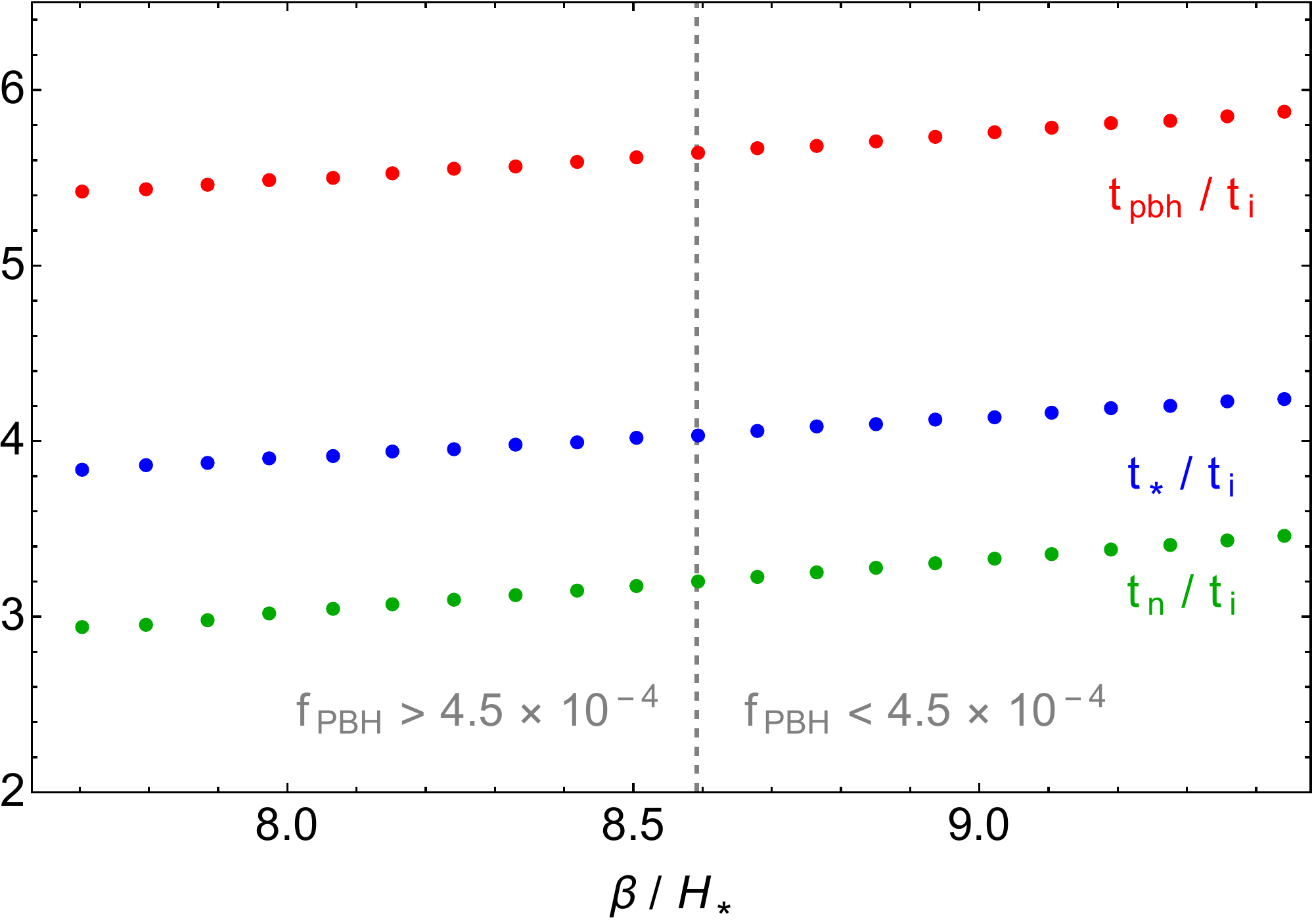}
	\includegraphics[width=0.44\textwidth]{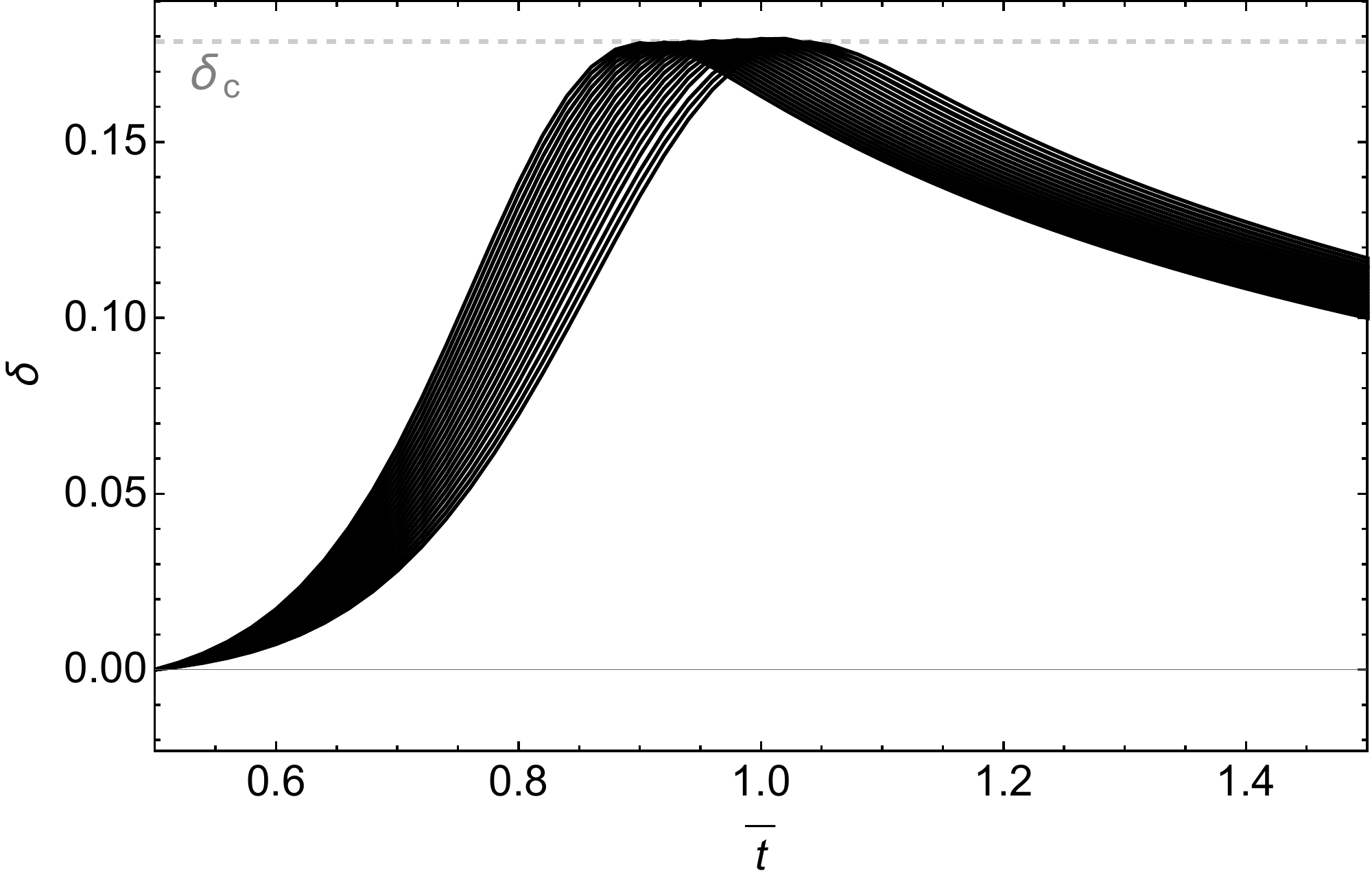}\\
	\includegraphics[width=0.46\textwidth]{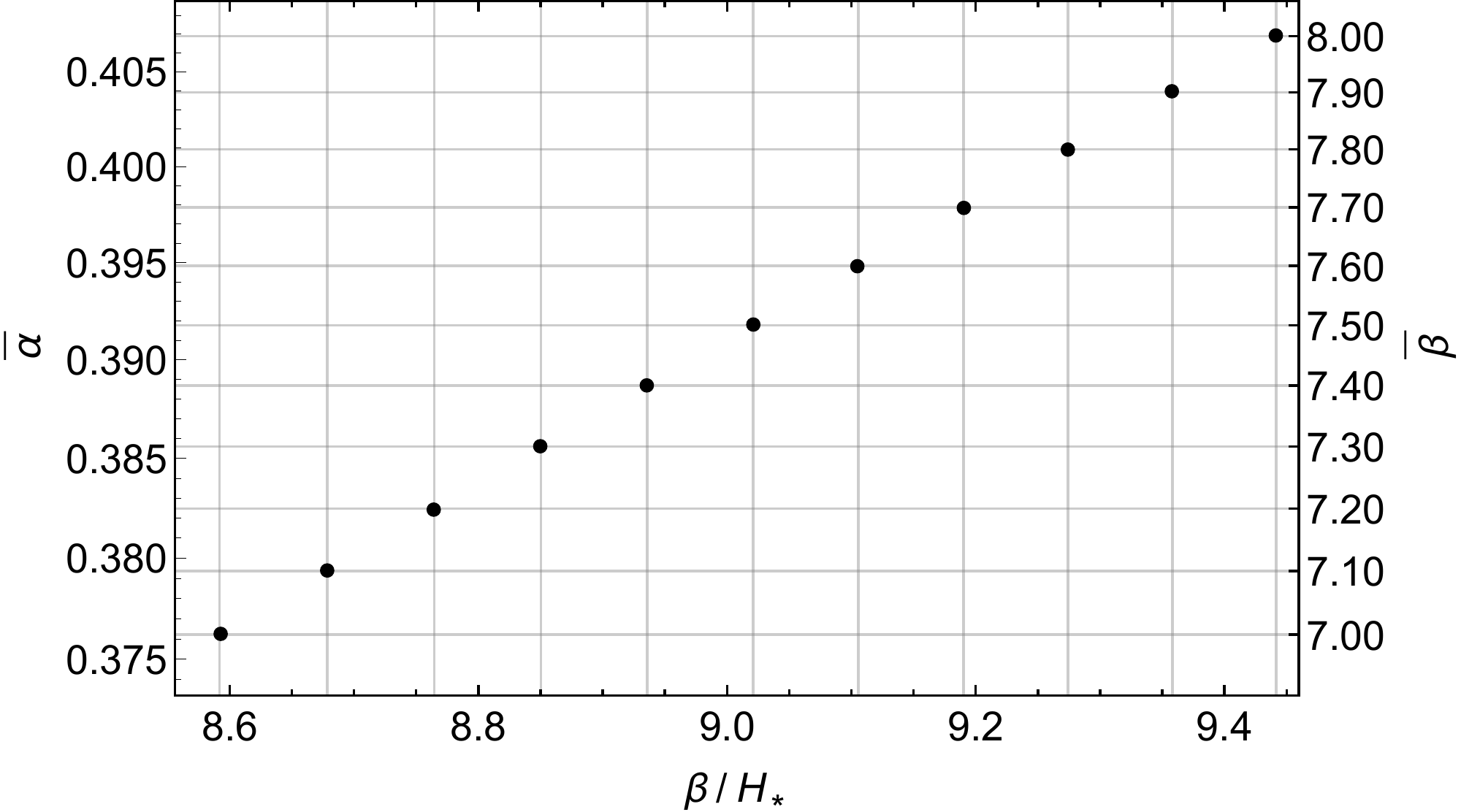}
	\includegraphics[width=0.45\textwidth]{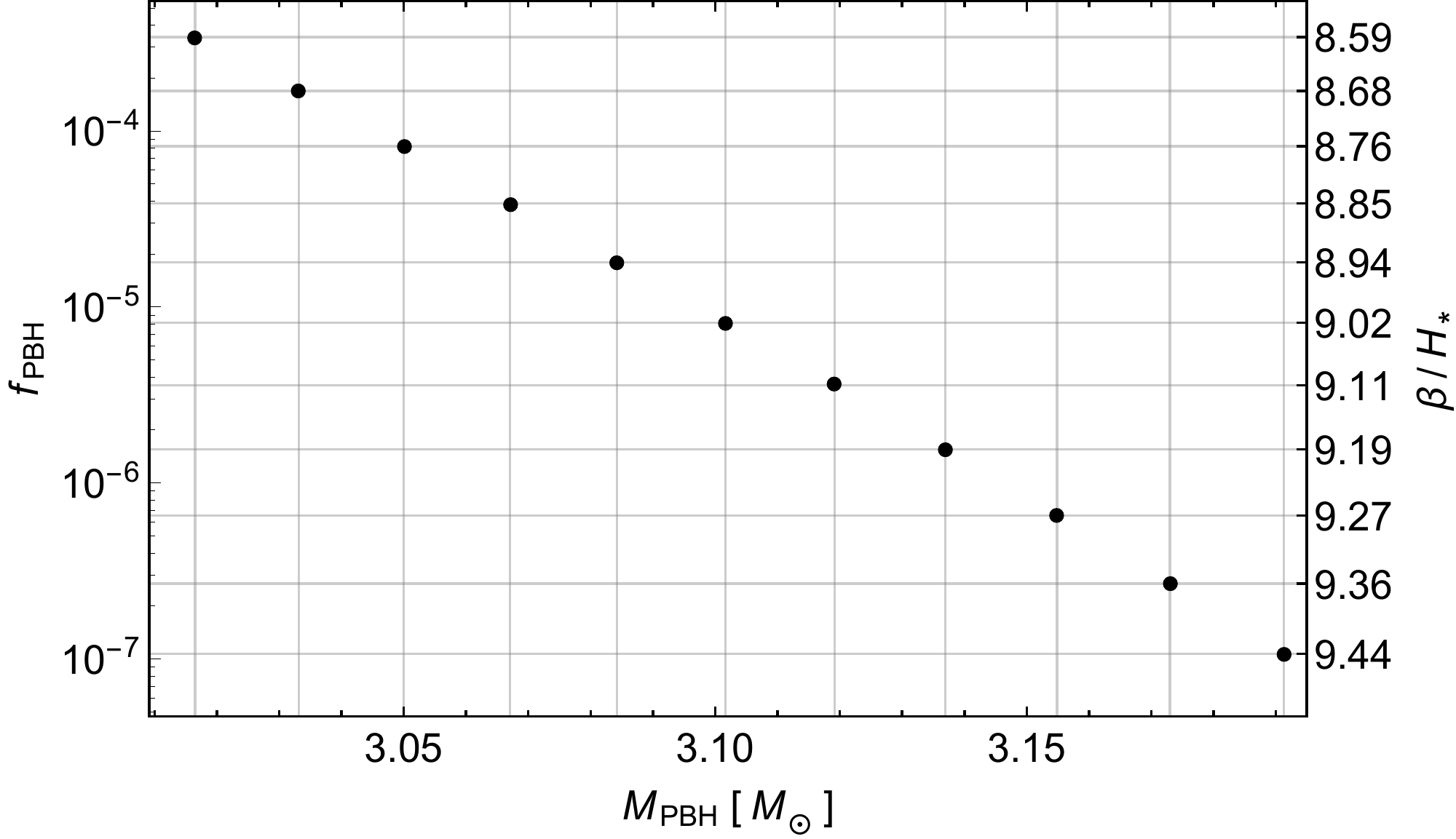}\\
	\caption{The results for the parameters in producing the SGWB and PBHs. \textit{Upper left}: The delayed decay time $t_n$ (green), the percolation time $t_*$ (blue), and the PBH formation time $t_\mathrm{PBH}$ (red) after normalized to the normal decay time $t_i$ with respect to the parameter regime of $\beta/H_*$ of interest.
		\textit{Upper right}: The time evolution of overdensity with respect to the parameter regime of $\beta/H_*$ of interest.
		\textit{Lower left}: The parameter space for $\bar{\alpha}$ and $\beta/H_*$ with respect to the input parameter $\bar{\beta}$.
		\textit{Lower right}: The parameter space for $f_\mathrm{PBH}$ and $M_\mathrm{PBH}$ with respect to the parameter regime of $\beta/H_*$ of interest.}
	\label{fig:PBHParameters}
\end{figure*}

We fill in some details for the PBH and SGWB productions associated with a FOPT and in particular for our holographic gluon model. The process of bubble nucleations and collisions mixed with PBH productions is highly inhomogeneous, therefore, a rigorous treatment would require for numerical simulations. A convenient approximation to the background evolution is close to the radiation-dominated era with $a(t)\propto t^{1/2}$, which will be checked and confirmed later as a good approximation for the parameter space we consider. We also normalize all dimensional quantities with the dimensional input $\Gamma_0^{1/4}$, such as $\bar{t}\equiv\Gamma_0^{1/4}t$, $\bar{\beta}\equiv\beta/\Gamma_0^{1/4}$, and $\bar{\alpha}\equiv\Delta V/(3M_\mathrm{Pl}^2\Gamma_0^{1/2})$.

During the asynochronized progress of PT, the fraction of spatial regions that are still staying at the false vacuum at time $t$ is now computed as
\begin{align}
F(\bar{t};\bar{t}_i,\bar{\beta})=\exp\left[-\frac43\pi\int_{\bar{t}_i}^{\bar{t}}\mathrm{d}\bar{t}'8e^{\bar{\beta}\bar{t}'}\left(\sqrt{\bar{t}\bar{t}'}-\bar{t}'\right)^3\right].
\end{align}
Without loss of generality, we can choose the true vacuum as the zero point of the potential energy so that the vacuum energy in the normal decay regions and delayed decayed regions are estimated by $\rho_v(\bar{t};\bar{t}_i)=F(\bar{t};\bar{t}_i)\Delta V$ and $\rho_v(\bar{t};\bar{t}_n)=F(\bar{t};\bar{t}_n)\Delta V$, respectively. The time evolution for the spatial fraction of false vacuum regions in the normal (solid curves) and delayed (dotted curves) decay channels is shown in the top panel of Fig.~\ref{fig:TimeEvolution}. For a larger value of $\bar{\beta}$, the PT proceeds more abruptly. For a smaller value of $\bar{\beta}$, the PT proceeds more slowly.

However, the rest of energy density does not evolve exactly as radiations due to the interrupt of the PT process and associated PBH productions. Nevertheless, the radiation evolution could be effectively solved from
\begin{align}\label{eq:normalizedEoM}
\frac{\mathrm{d}\bar{\rho}_r}{\mathrm{d}\bar{t}}+4\bar{\rho}_r\sqrt{\bar{\rho}_r+\bar{\rho}_v}=-\bar{\alpha}\frac{\mathrm{d}F}{\mathrm{d}\bar{t}},
\end{align}
with abbreviations $\bar{\rho}_r\equiv\rho_r/(3M_\mathrm{Pl}^2\Gamma_0^{1/2})$ and $\bar{\rho}_v\equiv\rho_v/(3M_\mathrm{Pl}^2\Gamma_0^{1/2})=F\bar{\alpha}$. Note that $\Gamma_0$ naturally defines a time $t_0$ in such a way that $H(t_0)=\Gamma_0^{1/4}$. Our assumption for radiation dominance requires $\bar{t}_0=H(t_0)t_0=1/2$. Then the initial condition is chose as $\bar{\rho}_r(\bar{t}_0)=1-\bar{\rho}_v(\bar{t}_0)$. For given $\bar{\alpha}$ and $\bar{\beta}$, Eq.~\ref{eq:normalizedEoM} can be solved for the normal and delayed decay channels respectively with $\bar{t}_n>\bar{t}_i$. It can be checked numerically that, as long as $\bar{\alpha}<0.5$, our assumption for the radiation dominance is valid throughout the whole process of PT as shown in the bottom left and bottom right panels of Fig.~\ref{fig:TimeEvolution} for the radiation/vacuum energy density fractions and radiation/vacuum energy densities normalized by the total energy density at the time $\bar{t}_0=1/2$, respectively. The normal decay time $\bar{t}_i=0.2$ and delayed decay time $\bar{t}_n=0.6$ as well as $\bar{\alpha}=0.3$ and $\bar{\beta}=7$ are chose for illustration.

With the full solutions for the radiation and vacuum energy densities in the normal and delayed decay regions, the overdensity of total energy density in the delayed decay regions can be directly evaluated by
\begin{align}
\delta(\bar{t})=\frac{\bar{\rho}_r(\bar{t}; \bar{t}_n)+\bar{F}(\bar{t}; \bar{t}_n)\bar{\alpha}}{\bar{\rho}_r(\bar{t}; \bar{t}_i)+\bar{F}(\bar{t}; \bar{t}_i)\bar{\alpha}}-1\,.
\end{align}
The time evolution of $\delta(\bar{t}; \bar{t}_i, \bar{t}_n, \bar{\alpha}, \bar{\beta})$ is first increasing due to the gradual accumulation of energy density in false vacuum and then decreasing due to the rapid declination of volume fraction in false vacuum. For given $\bar{t}_i$, $\bar{\alpha}$ and $\bar{\beta}$, when the maximal overdensity exactly saturates a given PBH threshold $\delta_c$, we can solve for the required delayed decay time $\bar{t}_n$, from which the PBH formation time is then solved from $\delta(\bar{t}_\mathrm{PBH}; \bar{t}_i, \bar{t}_n, \bar{\alpha}, \bar{\beta})=\delta_c$. Although suffered from large uncertainties of numerical simulations, we can adopt the analytic estimation~\cite{Harada:2013epa} on the PBH threshold via $\delta_c=\sin^2[\pi\sqrt{w}/(1+3w)]$ with the EoS $w$ evaluated from the dominant component at PBH formation.

The PBH mass produced from our postponed decay mechanism is almost monochromatic
since the numerical simulations for the gravitational collapse of over-dense regions with sub-horizon size are still missing. We therefore only focus on the PBH mass collapsed from the over-dense Hubble volumes with postponed decay,
\begin{align}
\frac{M_\mathrm{PBH}}{M_\odot}=4\pi\gamma_\mathrm{PBH}\left(\frac{M_\mathrm{Pl}}{M_\odot}\right)\left(\frac{M_\mathrm{Pl}/\Gamma_0^{1/4}}{H_\mathrm{PBH}/\Gamma_0^{1/4}}\right),
\end{align}
where $\gamma_\mathrm{PBH}=0.2$~\cite{Carr:1975qj}, $M_\mathrm{Pl}/M_\odot=2.182\times10^{-39}$, and $\bar{H}_\mathrm{PBH}\equiv H_\mathrm{PBH}/\Gamma_0^{1/4}=\sqrt{\bar{\rho}_\mathrm{tot}(\bar{t}_\mathrm{PBH};\bar{t}_n)}$. The other factor $M_\mathrm{Pl}/\Gamma_0^{1/4}$ can be roughly estimated as
\begin{align}
\frac{\Gamma_0^{1/4}}{M_\mathrm{Pl}}=\left(\frac{\pi^2}{90}g_\mathrm{dof}\right)^{1/2}\left(\frac{T_*}{M_\mathrm{Pl}}\right)^2e^{-\frac{\beta}{8H_*}}
\end{align}
by noting that the percolation time defined by $F(t_*;t_i)=0.7$ is usually close to the time when the bubble nucleation rate balances the Hubble expansion rate $\Gamma(t_*)\approx H(t_*)^4$, thus, $\Gamma_0\approx H_*^4e^{-\beta t_*}$ followed by the replacements of $3M_\mathrm{Pl}^2H_*^2=(\pi^2/30)g_\mathrm{dof}T_*^4$ and $H_*t_*=1/2$ due to the radiation dominance.

The PBH abundance $f_\mathrm{PBH}=(a_\mathrm{eq}/a_\mathrm{PBH})\Omega_\mathrm{PBH}/\Omega_\mathrm{DM}^\mathrm{eq}$ normalized to the dark matter fraction $\Omega_\mathrm{DM}^\mathrm{eq}=0.42$ at the matter-radiation equality is then estimated by
\begin{align}
\Omega_\mathrm{PBH}=\exp\left[-\frac43\pi\int_{\bar{t}_i}^{\bar{t}_n}\mathrm{d}\bar{t}\,e^{\bar{\beta}\bar{t}}\left(\frac{\sqrt{\bar{t}/\bar{t}_\mathrm{PBH}}}{\bar{H}_\mathrm{PBH}}\right)^3\right],
\end{align}
where the redshift factor $a_\mathrm{eq}/a_\mathrm{PBH}=T_\mathrm{PBH}/T_\mathrm{eq}$ with $T_\mathrm{eq}\approx0.75$ eV can be estimated by inserting $3M_\mathrm{Pl}^2H_\mathrm{PBH}^2=(\pi^2/30)g_\mathrm{dof}T_\mathrm{PBH}^4$ after replacing $H_\mathrm{PBH}=\bar{H}_\mathrm{PBH}\Gamma_0^{1/4}$ with previously computed $\bar{H}_\mathrm{PBH}$ and $\Gamma_0^{1/4}$, namely,
\begin{align}
T_\mathrm{PBH}=T_*\sqrt{\bar{H}_\mathrm{PBH}}e^{-\frac{\beta}{16H_*}}.
\end{align}
Note that the $g_\mathrm{dof}$-dependence in $T_\mathrm{PBH}$ and $\Gamma_0$ cancels out, leaving no dependence on $g_\mathrm{dof}$ for the PBH abundance. Finally, in computing both PBH mass and abundance, the inverse duration is determined by $\beta/H_*=\bar{\beta}/\sqrt{\bar{\rho}_\mathrm{tot}(\bar{t}_*;\bar{t}_i)}$.

For our holographic model of gluodynamics, $\beta$ is the only free parameter since one can further fix $\bar{\alpha}$ from matching
the strength factor $\alpha=\bar{\alpha}/\bar{\rho}_r(\bar{t}_*;\bar{t}_i)$ to the value $0.939$ obtained from holographic calculations. The other inputs from holographic calculations include the PT temperature $T_*=276.5$ MeV, the effective degrees of freedom $g_\mathrm{dof}=3.64$ and the PBH threshold $\delta_c=0.1786$ from the EoS $w\equiv P/\epsilon=0.0219$ of the dominant component in the unbroken phase~\cite{Harada:2013epa}.  The final results are summarized in Fig.~\ref{fig:PBHParameters}. In the first panel, all the characteristic time scales, such as the delayed decay time $t_n$, the percolation time $t_*$, and the PBH formation time $t_\mathrm{PBH}$, are shown with respect to $\beta/H_*$ after normalized to the normal decay time $t_i$. In the second panel, the time evolution of the overdensity within the delayed decay regions are shown to exactly saturate the PBH formation threshold. In the third and last panels, the parameter space for $\bar{\alpha}$, $\beta/H_*$, $f_\mathrm{PBH}$, and $M_\mathrm{PBH}$ are shown with respect to the input free parameter $\bar{\beta}$. It is worth noting that the current constraint from the current GWTC-3 data~\cite{Chen:2021nxo} on the PBH abundance $f_\mathrm{PBH}<0.00045$ in the mass range $[1\,M_\odot,3\,M_\odot]$ would constrain $\beta/H_*>8.59$.
\end{appendix}

\end{document}